\documentclass[sigplan, screen]{acmart}

\usepackage{cleveref}
\Crefname{algocf}{Algorithm}{Algorithms}
\usepackage{silence}
\usepackage{todonotes}
\usepackage{xspace}
\usepackage{subcaption}
\usepackage{listings}
\definecolor{lstpurple}{rgb}{0.5,0,0.5}
\definecolor{lstred}{rgb}{1,0,0}
\definecolor{lstreddark}{rgb}{0.7,0,0}
\definecolor{lstredl}{rgb}{0.64,0.08,0.08}
\definecolor{lstmildblue}{rgb}{0.66,0.72,0.78}
\definecolor{lstblue}{rgb}{0,0,1}
\definecolor{lstmildgreen}{rgb}{0.42,0.53,0.39}
\definecolor{lstgreen}{rgb}{0,0.5,0}
\definecolor{lstorangedark}{rgb}{0.6,0.3,0}
\definecolor{lstorange}{rgb}{0.75,0.52,0.005}
\definecolor{lstorangelight}{rgb}{0.89,0.81,0.67}
\definecolor{lstbeige}{rgb}{0.90,0.86,0.45}

\DeclareFontShape{OT1}{cmtt}{bx}{n}{<5><6><7><8><9><10><10.95><12><14.4><17.28><20.74><24.88>cmttb10}{}

\lstdefinestyle{psql}
{
tabsize=2,
basicstyle=\small\upshape\ttfamily,
language=SQL,
morekeywords={PROVENANCE,BASERELATION,INFLUENCE,COPY,ON,TRANSPROV,TRANSSQL,TRANSXML,CONTRIBUTION,COMPLETE,TRANSITIVE,NONTRANSITIVE,EXPLAIN,SQLTEXT,GRAPH,IS,ANNOT,THIS,XSLT,MAPPROV,cxpath,OF,TRANSACTION,SERIALIZABLE,COMMITTED,INSERT,INTO,WITH,SCN,UPDATED},
extendedchars=false,
keywordstyle=\bfseries,
mathescape=true,
escapechar=@,
sensitive=true
}

\lstdefinestyle{psqlcolor}
{
tabsize=2,
basicstyle=\small\upshape\ttfamily,
language=SQL,
morekeywords={PROVENANCE,BASERELATION,INFLUENCE,COPY,ON,TRANSPROV,TRANSSQL,TRANSXML,CONTRIBUTION,COMPLETE,TRANSITIVE,NONTRANSITIVE,EXPLAIN,SQLTEXT,GRAPH,IS,ANNOT,THIS,XSLT,MAPPROV,cxpath,OF,TRANSACTION,SERIALIZABLE,COMMITTED,INSERT,INTO,WITH,SCN,UPDATED},
extendedchars=false,
keywordstyle=\bfseries\color{lstblue},
deletekeywords={count,min,max,avg,sum},
keywords=[2]{count,min,max,avg,sum},
keywordstyle=[2]\color{lstblue},
stringstyle=\color{lstreddark},
commentstyle=\color{lstgreen},
mathescape=true,
escapechar=@,
sensitive=true
}

\lstdefinestyle{datalog}
{
basicstyle=\footnotesize\upshape\ttfamily,
language=prolog
}

\lstdefinestyle{pseudocode}
{
  tabsize=3,
  basicstyle=\small,
  language=c,
  morekeywords={if,else,foreach,case,return,in,or},
  extendedchars=true,
  mathescape=true,
  literate={:=}{{$\gets$}}1 {<=}{{$\leq$}}1 {!=}{{$\neq$}}1 {append}{{$\listconcat$}}1 {calP}{{$\cal P$}}{2},
  keywordstyle=\color{lstpurple},
  escapechar=&,
  numbers=left,
  numberstyle=\color{lstgreen}\small\bfseries,
  stepnumber=1,
  numbersep=5pt,
}

\lstdefinestyle{xmlstyle}
{
  tabsize=3,
  basicstyle=\small\upshape\ttfamily,
  language=xml,
  extendedchars=true,
  mathescape=true,
  escapechar=£,
  tagstyle=\bfseries,
  usekeywordsintag=true,
  morekeywords={alias,name,id},
  keywordstyle=\color{lstred}
}

\lstdefinestyle{xmlstyle-color}
{
  tabsize=3,
  basicstyle=\small\upshape\ttfamily,
  language=xml,
  extendedchars=true,
  mathescape=true,
  escapechar=£,
  tagstyle=\color{keywordpurple},
  usekeywordsintag=true,
  morekeywords={alias,name,id},
  keywordstyle=\color{lstred}
}

\usepackage{xspace}
\usepackage{float}
\usepackage[utf8]{inputenc}
\usepackage[linesnumbered,ruled,vlined]{algorithm2e}
\SetKwInput{KwInput}{Input}                
\SetKwInput{KwOutput}{Output}              
\usepackage{mathtools}
\DeclarePairedDelimiter{\ceil}{\lceil}{\rceil}
\usepackage{ifthen}

\newboolean{IsTechReport}
\setboolean{IsTechReport}{true}
\newcommand{\IfTechReport}[2]{\ifthenelse{\boolean{IsTechReport}}{#1}{#2}}
\newenvironment{sketch}{\noindent\emph{Proof Sketch.}}{}

\newcommand{\systemname}{\textsc{TreeToaster}\xspace}
\newcommand{\jitd}{\textsc{JustInTimeData}\xspace}
\newcommand{\dbt}{\textsc{DBToaster}\xspace}

\newcommand{\tinysection}[1]{\smallskip \noindent \textbf{#1.}~}

\newcommand{\tuple}[1]{\left<\;#1\;\right>}
\newcommand{\setof}[1]{\left\{\;#1\;\right\}}
\newcommand{\gmsetof}[1]{\left\{\hspace{-0.5mm}\left|\;#1\;\right|\hspace{-0.5mm}\right\}}
\newcommand{\comprehension}[2]{\left\{\;#1\;|\;#2\;\right\}}
\newcommand{\gmcomprehension}[2]{\gmsetof{\;#1\;|\;#2}}
\newcommand{\gOR}{\;|\;}
\newcommand{\isDefinedAs}{\overset{\vartriangle}{=}}
\newcommand{\True}{\mathbb{T}}
\newcommand{\False}{\mathbb{F}}
\newcommand{\trimfigurespacing}{}
\newcommand{\revised}[1]{\textcolor[HTML]{000000}{#1}}
\newcommand{\rerevised}[1]{\textcolor[HTML]{000000}{#1}}

\newcommand{\Schema}{\fancy{S}}
\newcommand{\Node}{N}
\newcommand{\Pattern}{q}
\newcommand{\Atom}{a}
\newcommand{\Constraint}{\theta}
\newcommand{\NodeExplicit}[3]{{(#1,#2,#3)}}
\newcommand{\PatternNodeName}{\texttt{Match}\xspace}
\newcommand{\PatternNode}[4]{\texttt{Match}(#1, #2, #3, #4)}
\newcommand{\Generate}{g}
\newcommand{\GenerateNodeName}{\texttt{Gen}\xspace}
\newcommand{\GenerateNode}[3]{\texttt{Gen}(#1, #2, #3)}
\newcommand{\GenerateExistingName}{\texttt{Reuse}\xspace}
\newcommand{\GenerateExisting}[1]{\texttt{Reuse}(#1)}
\newcommand{\Scopefunction}{\Gamma}

\newcommand{\Descendants}[1]{\texttt{Desc}(#1)}
\newcommand{\AnyNode}{\texttt{AnyNode}}
\newcommand{\Label}{\ell}
\newcommand{\AnnotationMap}{A}

\newcommand{\AnnotationVariable}{x}

\newcommand{\NodeVariable}{i}
\newcommand{\NodeChildren}{\overline{N}}
\newcommand{\PatternChildren}{\overline{Q}}
\newcommand{\emptyscope}{\emptyset}

\newcommand{\IVM}{\texttt{IVM}}

\newcommand{\InlineGenerate}{\texttt{Inline}_{gen}}

\newcommand{\IsLeaf}[1]{\textbf{isleaf}(#1)}
\newcommand{\DoGenerate}[1]{\llbracket #1\rrbracket}
\newcommand{\MatchPairs}[2]{\texttt{pair}(#1, #2)}
\newcommand{\GeneratePairs}[1]{\texttt{pair}(#1)}
\newcommand{\Ancestor}[2]{\texttt{Ancestor}_{#1}(#2)}

\newcommand{\NodeDomain}{\fancy{N}}

\newcommand{\LabelDomain}{\fancy{L}}
\newcommand{\ValueDomain}{\mathbb{D}}
\newcommand{\AnnotationVariableDomain}{\Sigma_{M}}
\newcommand{\NodeVariableDomain}{\Sigma_{\fancy{I}}}

\newcommand{\PatternDomain}{\fancy{Q}}

\newcommand{\View}[1]{\texttt{View}_{#1}}
\newcommand{\AtomDomain}{\texttt{atom}}
\newcommand{\ConstantDomain}{\texttt{const}}
\newcommand{\VariableDomain}{\NodeVariableDomain.\AnnotationVariableDomain}
\newcommand{\ConstraintDomain}{\Theta}
\newcommand{\Naturals}{\mathbb N}
\newcommand{\GenerateDomain}{\fancy{G}}

\newcommand{\NodeList}{[\Node_1 \ldots \Node_n]}

\newcommand{\PatternList}{[\Pattern_1 \ldots \Pattern_n]}

\newcommand{\Patternroot}{R}
\newcommand{\ASTroot}{\Node}

\newcommand{\Patterndepth}[1]{D(#1)}

\AtBeginDocument{%
  \providecommand\BibTeX{{%
    \normalfont B\kern-0.5em{\scshape i\kern-0.25em b}\kern-0.8em\TeX}}}

\setcopyright{none}

\usepackage{natbib}
\usepackage{graphicx}
\usepackage{amsthm}
\usepackage{mathpartir}
\usepackage{listings}
\usepackage{dsfont}
\newtheorem{definition}{Definition}
\newcommand{\fancy}[1]{{\mathcal{#1}}}
\theoremstyle{plain}

\definecolor{lstpurple}{rgb}{0.5,0,0.5}
\definecolor{lstred}{rgb}{1,0,0}
\definecolor{lstreddark}{rgb}{0.7,0,0}
\definecolor{lstredl}{rgb}{0.64,0.08,0.08}
\definecolor{lstmildblue}{rgb}{0.66,0.72,0.78}
\definecolor{lstblue}{rgb}{0,0,1}
\definecolor{lstmildgreen}{rgb}{0.42,0.53,0.39}
\definecolor{lstgreen}{rgb}{0,0.5,0}
\definecolor{lstorangedark}{rgb}{0.6,0.3,0}
\definecolor{lstorange}{rgb}{0.75,0.52,0.005}
\definecolor{lstorangelight}{rgb}{0.89,0.81,0.67}
\definecolor{lstbeige}{rgb}{0.90,0.86,0.45}

\DeclareFontShape{OT1}{cmtt}{bx}{n}{<5><6><7><8><9><10><10.95><12><14.4><17.28><20.74><24.88>cmttb10}{}

\lstdefinestyle{psqlcolor}
{
tabsize=2,
basicstyle=\small\upshape\ttfamily,
language=SQL,
morekeywords={PROVENANCE,BASERELATION,INFLUENCE,COPY,ON,TRANSPROV,TRANSSQL,TRANSXML,CONTRIBUTION,COMPLETE,TRANSITIVE,NONTRANSITIVE,EXPLAIN,SQLTEXT,GRAPH,IS,ANNOT,THIS,XSLT,MAPPROV,cxpath,OF,TRANSACTION,SERIALIZABLE,COMMITTED,INSERT,INTO,WITH,SCN,UPDATED},
extendedchars=false,
keywordstyle=\bfseries\color{lstblue},
deletekeywords={count,min,max,avg,sum},
keywords=[2]{count,min,max,avg,sum},
keywordstyle=[2]\color{lstblue},
stringstyle=\color{lstreddark},
commentstyle=\color{lstgreen},
mathescape=true,
escapechar=@,
sensitive=true
}

\lstdefinestyle{scala}
{
  tabsize=3,
  basicstyle=\footnotesize\ttfamily,
  language=scala,
  morekeywords={if,else,foreach,case,return,in,or,@,_},
  extendedchars=true,
  mathescape=true,
  literate={:=}{{$\gets$}}1 {<=}{{$\leq$}}1 {!=}{{$\neq$}}1 {append}{{$\listconcat$}}1 {calP}{{$\cal P$}}{2},
  keywordstyle=\color{lstpurple},
  stepnumber=1,
  numbersep=5pt,
}

\lstdefinestyle{cpp}
{
  tabsize=3,
  basicstyle=\scriptsize\ttfamily,
  language=c,
  morekeywords={if,else,foreach,case,return,in,or},
  extendedchars=true,
  mathescape=true,
  literate={:=}{{$\gets$}}1 {<=}{{$\leq$}}1 {!=}{{$\neq$}}1 {append}{{$\listconcat$}}1 {calP}{{$\cal P$}}{2},
  keywordstyle=\color{lstpurple},
  stepnumber=1,
  numbersep=5pt,
}

\begin{document}
\sloppy

\title{\systemname: Towards an IVM-Optimized Compiler}


\author{Darshana Balakrishnan, Carl Nuessle, Oliver Kennedy, Lukasz Ziarek}
\affiliation{%
  \institution{University at Buffalo}
}
\email{[ dbalakri, carlnues,okennedy, lziarek ]@buffalo.edu}

\renewcommand{\shortauthors}{D.Balakrishnan, et al.}

\begin{abstract}
  A compiler's optimizer operates over abstract syntax trees (ASTs), continuously applying rewrite rules to replace subtrees of the AST with more efficient ones.
  Especially on large source repositories, even simply finding opportunities for a rewrite can be expensive, as optimizer traverses the AST naively.
  In this paper, we leverage the need to repeatedly find rewrites, and explore options for making the search faster through indexing and incremental view maintenance (IVM).
  Concretely, we consider bolt-on approaches that make use of embedded IVM systems like DBToaster, as well as two new approaches: Label-indexing and \systemname, an AST-specialized form of IVM.
  We integrate these approaches into an existing just-in-time data structure compiler and show experimentally that \systemname can significantly improve performance with minimal memory overheads.
\end{abstract}


\maketitle

\section{Introduction}
\label{sec:introduction}

\begin{figure}[tb]
  \centering
  \includegraphics[width=\columnwidth]{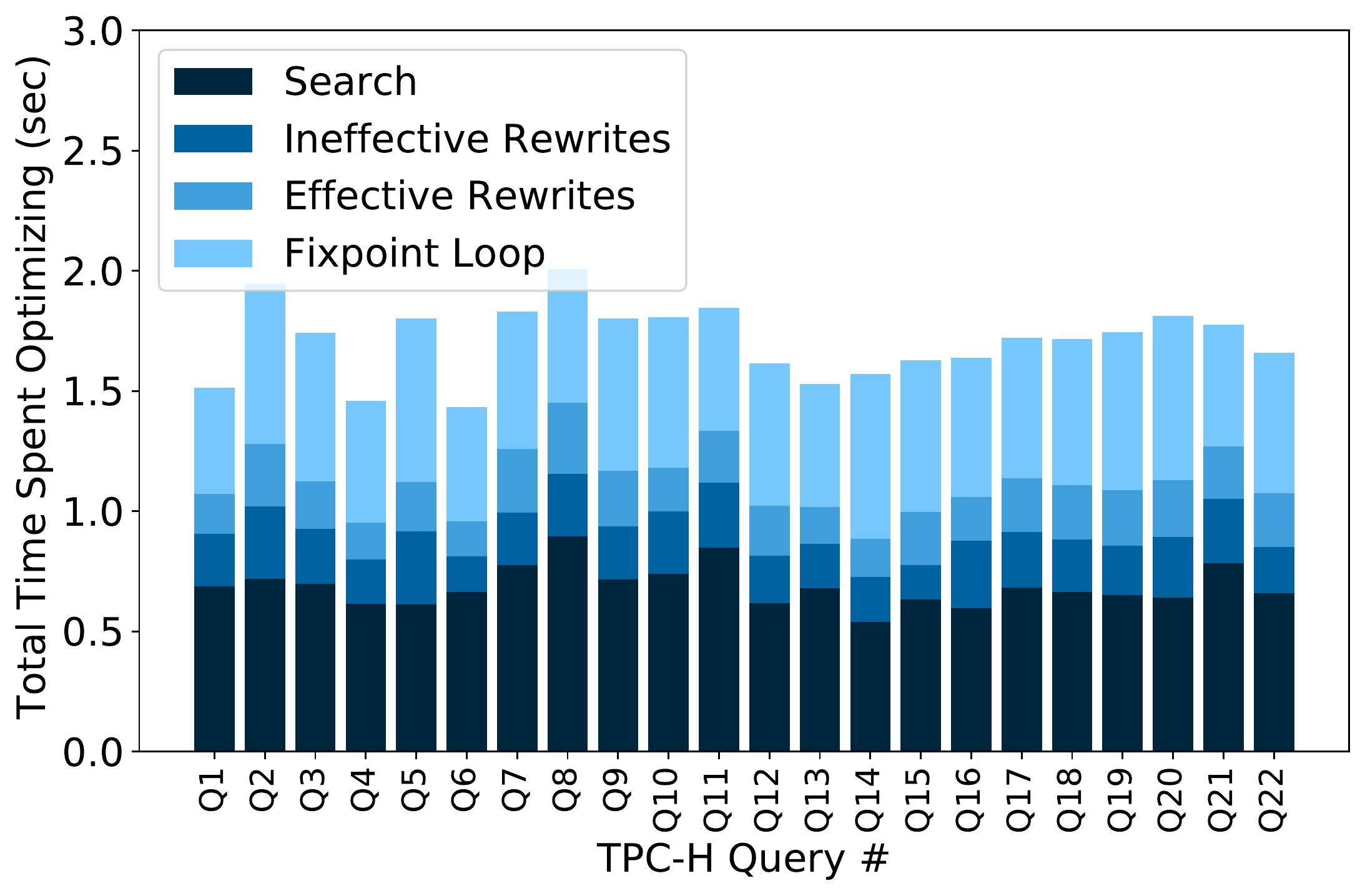}
  \caption{\rerevised{
    Time breakdown of the Catalyst optimizer on the 22 queries of the TPC-H Benchmark, including
    (i) searching the AST for candidate rewrites (\textbf{Search}), (ii) constructing new AST subtrees before aborting (\textbf{Ineffective}), (iii) constructing new AST subtrees (\textbf{Effective}), and (iv) In the optimizer's outer loop looking for a fixpoint (\textbf{Fixpoint}).
  }}
  \label{fig:exampleSearchTime}
  \trimfigurespacing
\end{figure}

\revised{
Typical database query optimizers, like Apache Spark's Catalyst~\cite{DBLP:conf/sigmod/ArmbrustXLHLBMK15} and Greenplum's Orca~\cite{DBLP:conf/sigmod/SolimanAREGSCGRPWNKB14}, work with queries encoded as abstract syntax trees (ASTs).
A tree-based encoding makes it possible to specify optimizations as simple, composable, easy-to-reason-about pattern/replacement rules.
}
\rerevised{
However, such pattern matching can be very slow.
For example, \Cref{fig:exampleSearchTime} shows a breakdown of how Catalyst spends its time optimizing \footnote{The instrumented version of spark can be found at \url{https://github.com/UBOdin/spark-instrumented-optimizer}} the 22 queries of the TPC-H benchmark~\cite{tpch}.  
33-45\% of its time is spent searching for optimization opportunities, using Scala's \texttt{match} operator to recursively pattern-match with every node of the tree
A further 27-43\% of the optimizer's time is spent in optimizer's outer fixpoint loop (e.g., comparing ASTs to decide whether the optimizer has converged or if further optimization opportunities might exist).
}
\revised{
On larger queries, pattern matching can be tens or even hundreds of seconds\footnote{
This number is smaller, but still notable for Orca, accounting for 5-20\% of the optimizer's time on a similar test workload.
}.
}

In this paper, we propose \systemname, an approach to incremental view maintenance specialized for use in compilers.  
As we show, \systemname virtually eliminates the cost of finding nodes eligible for a rewrite.
In lieu of repeated linear scans through the AST for eligible nodes, \systemname materializes a view for each rewrite rule, containing all nodes eligible for the rule, and incrementally maintains it as the tree evolves through the optimizer.

\begin{figure}[tb]
  \centering
  \includegraphics[width=0.9\columnwidth]{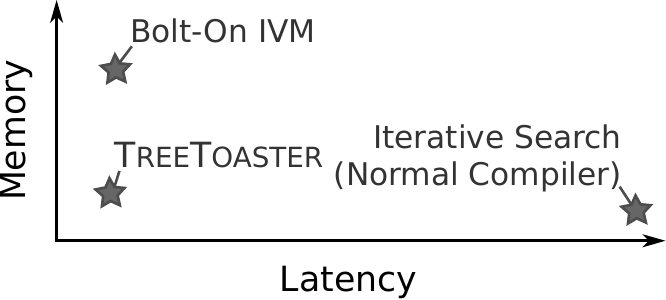}
  \caption{\systemname achieves AST pattern-matching performance competitive with ``bolting-on'' an embedded IVM system, but with negligible memory overhead.}
  \label{fig:overview}
\end{figure}

\revised{
Naively, we might implement this incremental maintenance scheme by simply reducing the compiler's pattern matching logic to a standard relational query language, and ``bolting on'' a standard database view maintenance system~\cite{DBLP:conf/sigmod/RossSS96,DBLP:journals/vldb/KochAKNNLS14}.
This simple approach typically reduces search costs to a (small) constant, while adding only a negligible overhead to tree updates.
However, classical view maintenance systems come with a significant storage overhead.
}
As we show in this paper, \systemname improves on the  ``bolt-on'' approach by leveraging the fact that both ASTs and pattern queries are given as trees.
As we show, when the data and query are both trees, \systemname achieves similar maintenance costs without the memory overhead of caching intermediate results (\Cref{fig:overview}).
\systemname further reduces memory overheads by taking advantage of the fact that the compiler already maintains a copy of the AST in memory, with pointers linking nodes together. 
\systemname combines these compiler-specific optimizations with standard techniques for view maintenance (e.g., inlining and compiling to C++~\cite{DBLP:journals/vldb/KochAKNNLS14}) to produce an incremental-view maintenance engine that meets or beats state-of-the-art view maintenance on AST pattern-matching workloads, while using much less memory.

\revised{
To illustrate the advantages of \systemname, we apply it to a recently proposed Just-in-Time Data Structure compiler~\cite{DBLP:conf/cidr/KennedyZ15,DBLP:conf/dbpl/BalakrishnanZK19} that reframes tree-based index data structures as ASTs.
Like other AST-based optimizers, pattern/replacement rules asynchronously identify  opportunities for incremental reorganization like database cracking~\cite{DBLP:conf/cidr/IdreosKM07} or log-structured merges~\cite{DBLP:journals/acta/ONeilCGO96}.
We implement \systemname within \jitd and show that it virtually eliminates AST search costs with minimal memory overhead.
}

\revised{
Concretely, the contributions of this paper are:
(i) We formally model AST pattern-matching queries and present a technique for incrementally maintaining precomputed views over such queries;
(ii) We show how declaratively specified rewrite rules can be further inlined into view maintenance to further reduce maintenance costs;
(iii) As a proof of concept, we ``bolt-on'' DBToaster, an embeddable IVM system, onto a just-in-time data-structure compiler~\cite{DBLP:conf/cidr/KennedyZ15,DBLP:conf/dbpl/BalakrishnanZK19}.  This modification dramatically improves performance, but adds significant memory overheads;}
\rerevised{(iv) We present \systemname, a IVM system optimized for compiler construction.  \systemname avoids the high memory overheads of bolt-on IVM;}
\revised{}{(v) We present experiments that show that \systemname significantly outperforms ``bolted-on'' state-of-the-art IVM systems and is beneficial to the \jitd compiler.
}

\section{Notation and Background}
\label{sec:background}
In its simplest form, a typical compiler's activities break down into three steps: parsing, optimizing, and output.

\tinysection{Parsing}
First, a parser converts input source code into a structured Abstract Syntax Tree (AST) encoding of the source.

\begin{figure}
\includegraphics[width=0.9\columnwidth]{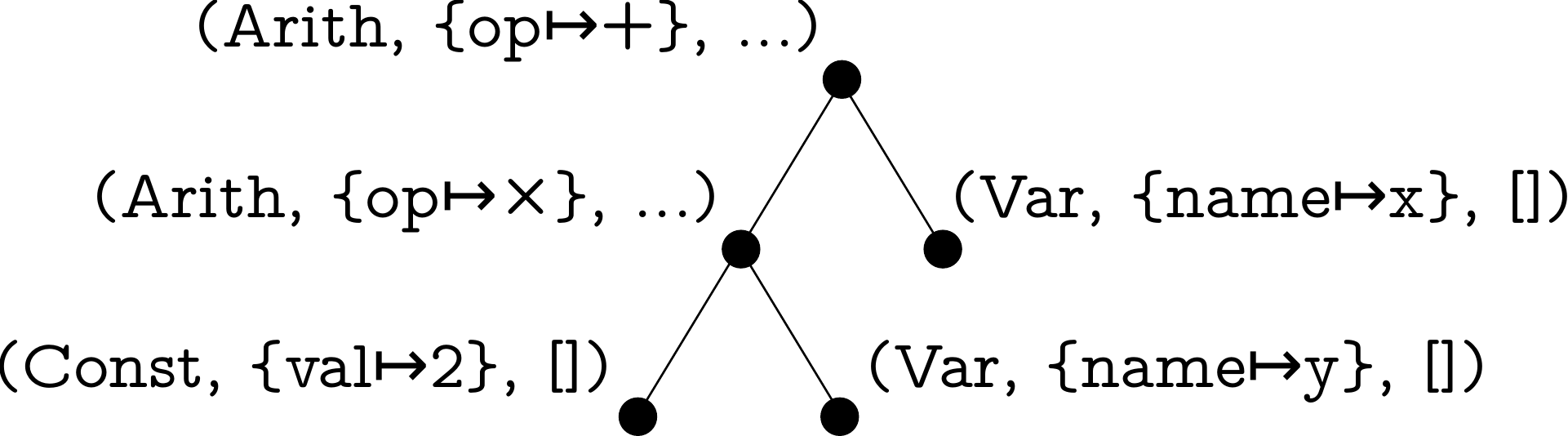}
\caption{An AST for the expression \lstinline{2 * y + x}}
\label{fig:exampleAST}
\Description{
  An example abstract syntax tree for the expression `x plus two times y'.  
  A variable `+' node is at the root with two children: an arithmetic times node and a variable x.  
  The arithmetic plus node has two children: a constant two node and a variable y node.
} 
\end{figure}

\begin{example}
\revised{
\Cref{fig:exampleAST} shows the AST for the expression \lstinline{2 * y + x}.
AST nodes have labels (e.g., \lstinline{Arith}, \lstinline{Var}, or \lstinline{Const}) and attributes (e.g., $\{\texttt{op} \mapsto +\}$ or $\{\texttt{val} \mapsto 2\}$).
}
\end{example}

We formalize an AST as a tree with labeled nodes, each annotated with zero or more attributes.

\begin{definition}[Node] 
An Abstract Syntax Tree node \revised{$\Node = \NodeExplicit{\Label}{\AnnotationMap}{\NodeChildren}$} is a 3-tuple, consisting of (i) a label $\Label$ drawn from an alphabet $\LabelDomain$; (ii) annotations $\AnnotationMap: \AnnotationVariableDomain \rightarrow \ValueDomain$, a partial map from an alphabet of attribute names $\AnnotationVariableDomain$ to a domain $\ValueDomain$ of attribute values; and (iii) an ordered list of children $\NodeChildren$.
\end{definition}

We define a leaf node (denoted $\IsLeaf\Node$) as a node that has no child nodes (i.e., $\NodeChildren = \emptyset$).
We assume that nodes follow a schema $\Schema : \LabelDomain \rightarrow 2^{\AnnotationVariableDomain} \times \Naturals$;
For each label ($\Label \in \Schema$), we fix a set of attributes that are present in all nodes with the label ($\overline{x} \in 2^{\AnnotationVariableDomain}$), as well as an upper bound on the number of children ($c \in \Naturals$).

\tinysection{Optimization}
Next, the optimizer rewrites the AST, iteratively applying pattern-matching expressions and deriving replacement subtrees.
\revised{
We note that even compilers written in imperative languages frequently adopt a declarative style for expressing pattern-matching conditions.
For example, ORCA~\cite{DBLP:conf/sigmod/SolimanAREGSCGRPWNKB14} (written in C++) builds small ASTs to describe pattern matching structures, while Catalyst~\cite{DBLP:conf/sigmod/ArmbrustXLHLBMK15} (written in Scala) relies on Scala's native pattern-matching syntax.
}

\begin{example}
\label{ex:inlineArithmetic}
A common rewrite rule for arithmetic eliminates no-ops like addition to zero.  For example, the subtree
\begin{align*}
(\texttt{Arith}, \{\texttt{op}\mapsto+\}, [\;\; &
  \NodeExplicit{\texttt{Const}}{\{\texttt{val}\mapsto 0\}}{[]},\\ &
  \NodeExplicit{\texttt{Var}}{\{\texttt{name}\mapsto b\}}{[]}
\;\;]) 
\end{align*}
can be replaced by $\NodeExplicit{\texttt{Var}}{\{\texttt{name}\mapsto b\}}{[]}$
If the optimizer encounters a subtree with an \lstinline{Arith} node at the root, \lstinline{Const} and \lstinline{Var} nodes as children, and a 0 as the \texttt{val} attribute of the \lstinline{Const} node; it replaces the entire subtree by the \lstinline{Var} node.
\end{example}

The optimizer continues searching for subtrees matching one of its patterns until no further matches exist (a fixed point), or an iteration threshold or timeout is reached.

\tinysection{Output}
Finally, the compiler uses the optimized AST as appropriate (e.g., by generating bytecode or a physical plan).

\subsection{Pattern Matching Queries}
We formalize pattern matching in the following grammar:

\begin{definition}[Pattern]
A pattern query $q\in \fancy{Q}$ is one of
\revised{
\begin{mathpar}
    \PatternDomain : \AnyNode \gOR \PatternNode{\LabelDomain}{\NodeVariableDomain}{\PatternChildren}{\ConstraintDomain}
\end{mathpar}
}
\end{definition}
\vspace*{-4mm}
The symbol $\PatternNode{\Label_q}{\NodeVariable}{\PatternChildren}{\Constraint}$ indicates a structural match that succeeds iff 
(i) The matched node has label $\Label_q$, 
(ii) the children of the matched node recursively satisfy $\Pattern_i \in \PatternChildren$, and 
(iii) the constraint $\Constraint$ over the attributes of the node and its children is satisfied.
The node variable $\NodeVariable \in \NodeVariableDomain$ is used to identify the node in subsequent use, for example to reference the node's attributes in the constraint ($\Constraint$). 
The symbol $\AnyNode$ matches any node.  
\Cref{fig:patternMatchRules} formalizes the semantics of the $\PatternDomain$ grammar.

The grammar for constraints is given in \Cref{fig:constraint grammar}, and its semantics are typical.
A variable atom $\NodeVariable.\AnnotationVariable$ is a 2-tuple of a Node name ($\NodeVariable \in \NodeVariableDomain$) and an Attribute name ($\AnnotationVariable \in \AnnotationVariableDomain$), respectively, and evaluates to $\Scopefunction(\NodeVariable)(\AnnotationVariable)$, given some scope $\Scopefunction : \NodeVariableDomain \rightarrow \AnnotationVariableDomain \rightarrow \ValueDomain$.
This grammar is expressive enough to capture the full range of comparisons ($>$, $\geq$, $\leq$, $<$, $=$, $<$), and so we use these freely throughout the rest of the paper.

\begin{figure}
{\small
$$\ConstraintDomain: \AtomDomain=\AtomDomain \gOR
                                  \AtomDomain<\AtomDomain \gOR 
                                  \ConstraintDomain \wedge \ConstraintDomain \gOR 
                                  \ConstraintDomain \vee \ConstraintDomain \gOR
                                  \neg \ConstraintDomain \gOR
                                  \True \gOR \False
$$
$$\AtomDomain: \ConstantDomain \gOR \VariableDomain \gOR \AtomDomain\; [+,-,\times,\div]\; \AtomDomain$$
}\vspace*{-6mm}
\caption{Constraint Grammar}
\label{fig:constraint grammar}
\Description{textual figure}
\end{figure}

\begin{figure}
\revised{\small
$$
\llbracket \Pattern(\Label,\AnnotationMap,\NodeList) \rrbracket = \begin{cases}
    \True, \emptyscope    & \textbf{if } \Pattern = \AnyNode\\
    \True, \Gamma       & \textbf{if } \Pattern =  \PatternNode{\Label_\Pattern}{\NodeVariable}{\PatternList}{\Constraint} \\
                        & \hspace{4mm}\Label_\Pattern =\Label, \hspace{4mm}\Constraint(\Scopefunction),\\
                        & \hspace{4mm}\llbracket\Pattern_1(\Node_1)\rrbracket = \True,\Scopefunction_1 \\  
                        & \hspace{4mm}\ldots\llbracket\Pattern_n(\Node_n)\rrbracket = \True,\Scopefunction_n, \\
                        & \hspace{4mm}\Scopefunction = \setof{\NodeVariable\rightarrow\AnnotationMap} \cup 
                         {\bigcup_{k \in [n]}} \Scopefunction_k\\
    \False, \emptyscope   & \textbf{otherwise}
\end{cases}$$
}
\caption{Semantics for pattern queries ($\Pattern \in \PatternDomain$)}
\label{fig:patternMatchRules}
\trimfigurespacing
\end{figure}

\begin{example}
\label{ex:matchArithmetic}
\revised{
Returning to \Cref{ex:inlineArithmetic}, only \lstinline{Arith} nodes over \lstinline{Const} and \lstinline{Var} nodes as children are eligible for the simplification rule.
The corresponding pattern query is:
\begin{align*}
\PatternNode{\texttt{Arith}}{A}{[ \;&
  \PatternNode{\texttt{Const}}{B}{[]}{\{B.\texttt{val} = 0\}},\\ &
  \PatternNode{\texttt{Var}}{C}{[]}{\True}
\;]}{\{A.\texttt{op} = +\}}
\end{align*}
Note the constraint on the \lstinline{Const} match pattern; This subpattern only matches a node with a \texttt{val}(ue) attribute of $0$.
}
\end{example}

We next formalize pattern matching over ASTs.  First, we define the descendants of a node (denoted $\Descendants{\Node}$) to be the set consisting of $\Node$ and its descendants:
{\small
$$\Descendants{\Node} \isDefinedAs \setof{\Node} \bigcup_{k\in[n]} \Descendants{\Node_k}
\text{ s.t. } \Node = \NodeExplicit{\Label}{\AnnotationMap}{[\Node_1, \ldots, \Node_n]}$$
}
\begin{definition}[Match]
A match result, denoted $\Pattern(\Node)$, is the subset of $\Node$ or its descendents on which $\Pattern$ evaluates to true.
$$\Pattern(\Node) \isDefinedAs
\comprehension{\Node'}{\Node' \in \Descendants{\Node} \wedge \exists \Scopefunction : \Pattern(\Node') = \True, \Scopefunction}
$$

\end{definition}

\tinysection{Pattern Matching is Expensive}
Optimization is a tight loop in which the optimizer searches for a pattern match, applies the corresponding rewrite rule to the matched node, and repeats until convergence.
Pattern matching typically requires iteratively traversing the entire AST.
Every applied rewrite creates or removes opportunities for further rewrites, necessitating repeated searches for the same pattern.
Even with intelligent scheduling of rewrites, the need for repeated searches can not usually be eliminated outright, and as shown in \Cref{fig:exampleSearchTime} can take up to 45\% of the optimizer's time.

\begin{example}
Continuing the example, the optimizer would traverse the entire AST looking for \lstinline{Arith} nodes with the appropriate child nodes.
A depth-first traversal ensures that any replacement happens before the optimizer checks the parent for eligibility.
However, another rewrite may introduce new opportunities for simplification (e.g., by creating new \lstinline{Const} nodes), and the tree traversal must be repeated.
\end{example}

\section{Bolting-On IVM for Pattern Matching}
\label{sec:bolton}
\revised{
As a warm-up, we start with a simple, naive implementation of incremental view maintenance for compilers by mapping our pattern matching grammar onto relational queries, and ``bolting on'' an existing system for incremental view maintenance (IVM).
Although this specific approach falls short, it illustrates how IVM relates to the pattern search problem.
}
To map the AST to a relational encoding, for each label/schema pair $\Label \rightarrow \tuple{\setof{\AnnotationVariable_1, \ldots, \AnnotationVariable_k}, c} \in \Schema$, we define a relation $R_{\Label}(\texttt{id}, \AnnotationVariable_1, \ldots, \AnnotationVariable_k, \texttt{child}_{1}, \ldots, \texttt{child}_c)$ with an id field, and one field per attribute or child.
Each node $\Node = \NodeExplicit{\Label}{\AnnotationMap}{[\Node_1, \ldots, \Node_c]}$ is assigned a unique identifier $\textbf{id}_\Node$ and defines a row of relation $R_\Label$.
$$\tuple{\textbf{id}_\Node, \AnnotationMap(\AnnotationVariable_1), \ldots, \AnnotationMap(\AnnotationVariable_k), \textbf{id}_{\Node_1}, \ldots, \textbf{id}_{\Node_c}}$$

\begin{figure}
{\small
$$\overline R_\Pattern \isDefinedAs \begin{cases}
  \emptyset &\textbf{if } \Pattern = \AnyNode\\
  \setof{(R_\Label \texttt{ AS } \NodeVariable)} \underset{x \in [n]}{\bigcup} \overline R_{\Pattern_x} &\textbf{if } \Pattern = \PatternNode{\Label}{\NodeVariable}{[\Pattern_1, \ldots, \Pattern_n]}{\Constraint}
\end{cases}$$
}
{\small
$$\Constraint_\Pattern \isDefinedAs \begin{cases}
  \True &\textbf{if } \Pattern = \AnyNode\\
  \Constraint \underset{x \in [n]}{\bigwedge} \Constraint_{\Pattern_x} \wedge \texttt{join}(i.\texttt{child}_x, \Pattern_x) \hspace*{-10mm}\vspace*{-3mm}\\

    &\textbf{if } \Pattern = \PatternNode{\Label}{\NodeVariable}{[\Pattern_1, \ldots, \Pattern_n]}{\Constraint}
\end{cases}$$
}
{\small
$$\texttt{join}(\Atom, \Pattern) \isDefinedAs \begin{cases}
  \True                  &\textbf{if } \Pattern = \AnyNode\\
  \Atom = i.\textbf{id}  &\textbf{if } \Pattern = \PatternNode{\Label}{\NodeVariable}{[\Pattern_1, \ldots, \Pattern_n]}{\Constraint}
\end{cases}$$
}
\begin{lstlisting}
      $\Pattern \equiv$ SELECT * FROM $\overline R_\Pattern$ WHERE $\Constraint_\Pattern$
\end{lstlisting}
\caption{Converting a pattern $\Pattern$ to an equivalent SQL query.}
\label{fig:sqlQueryForPattern}
\trimfigurespacing
\end{figure}

A pattern $\Pattern$ can be reduced to an equivalent query over the relational encoding, as shown in \Cref{fig:sqlQueryForPattern}.  
A pattern with $k$ \PatternNodeName nodes becomes a $k$-ary join over the relations $\overline R_\Pattern$ corresponding to the label on each \PatternNodeName node.
Each relation is aliased to its node variable.
Join constraints are given by parent/child relationships, and pattern constraints transfer directly to the \lstinline{WHERE} clause.

\begin{example}
\revised{
Continuing \Cref{ex:inlineArithmetic}, the AST nodes are encoded as relations:
\texttt{Arith(id, op, child$_1$, child$_2$)}, \texttt{Const(id, val)}, and 
\texttt{Var(id, name)}.  The corresponding pattern match query, following the process in \Cref{fig:sqlQueryForPattern} is:
}{
\begin{lstlisting}
SELECT * FROM Arith a, Const b, Var c
WHERE a.child$_1$ = b.id AND a.child$_2$ = c.id
  AND a.op = '+' AND b.val = 0
\end{lstlisting}
}
\end{example}


\subsection{Background: Incremental View Maintenance}
Materialized views are used in production databases to accelerate query processing.
If a view is accessed repeatedly, database systems materialize the view query $Q$ by precomputing its results $Q(D)$ on the database $D$.
When the database changes, the view must be updated to match:
Given a set of changes, $\Delta D$ (e.g., insertions or deletions), a naive approach would be to simply recompute the view on the updated database $Q(D + \Delta D)$.
However, if $\Delta D$ is small, most of this computation will be redundant.
A more efficient approach is to derive a so-called ``delta query'' ($\Delta Q$) that computes a set of updates to the (already available) $Q(D)$. That is, denoting the view update operation by $\Leftarrow$:
$$Q(D+\Delta D) \;\;\;\; \equiv \;\;\;\; Q(D) \Leftarrow \Delta Q(D, \Delta D)$$

\begin{example}
\revised{
Recall $Q(\texttt{Arith}, \texttt{Const}, \texttt{Var})$ from the prior example.
After inserting a row $c$ into $\texttt{Const}$, we want:\\
$Q(\texttt{Arithmetic}, \texttt{Const} \uplus c, \texttt{Var})$
\vspace*{-1mm}\begin{align*}
\hspace{5mm}&=& \texttt{Arith} \bowtie (\texttt{Const} \uplus c) \bowtie \texttt{Var}\\
&=& (\texttt{Arith} \bowtie \texttt{Const} \bowtie \texttt{Var})
   \uplus (\texttt{Arith} \bowtie c \bowtie \texttt{Var})\\
&=& Q(\texttt{Arith}, \texttt{Const}, \texttt{Var}) \uplus (\texttt{Arith} \bowtie c \bowtie \texttt{Var})
\end{align*}
Instead of computing the full 3-way join, we can replace $c$ with a singleton and compute the cheaper query $(\texttt{Arith} \bowtie c \bowtie \texttt{Var})$, and union the result with our original materialzed view to obtain an updated view.
}
\end{example}

The joint cost of $\Delta Q(D, \Delta D)$ and the $\Leftarrow$ operator is generally lower than re-running the query, significantly improving overall performance when database updates are small and infrequent.  
However, $\Delta Q$ can still be expensive.
For larger or more frequent changes, we can further reduce the cost of computing $\Delta Q$ by caching intermediate results.
Ross et. al.~\cite{DBLP:conf/sigmod/RossSS96} proposed a form of cascading IVM that caches every intermediate result relation in the physical plan of the view query.

\begin{example}
\revised{Consider the running example query with the following execution order:
$$(\texttt{Arithmetic} \bowtie \texttt{Var}) \bowtie \texttt{Const}$$
In addition to materializing $Q(\cdot)$, Ross' scheme also materializes the results of $Q_1 = (\texttt{Arithmetic} \bowtie \texttt{Var})$.
When $c$ is inserted into $\texttt{Const}$, 
}
the update only requires a simple 2-way join $c \bowtie Q_1$.
However, updates are now (slightly) more expensive as multiple views may need to be updated.
\end{example}

Ross' approach of caching intermediate state is analogous to typical approaches to fixpoint computation (e.g., Differential Dataflow~\cite{DBLP:conf/cidr/McSherryMII13}), but penalizes updates to tables early in the query plan.  
With DBToaster~\cite{DBLP:journals/vldb/KochAKNNLS14}, Koch et. al. proposed instead materializing all possible query plans.
Counter-intuitively, this added materialization significantly reduces the cost of view maintenance. 
Although far more tables need to be updated with every database change, the updates are generally small and efficiently computable.

\begin{figure}
  \centering
  \includegraphics[width=0.8\columnwidth]{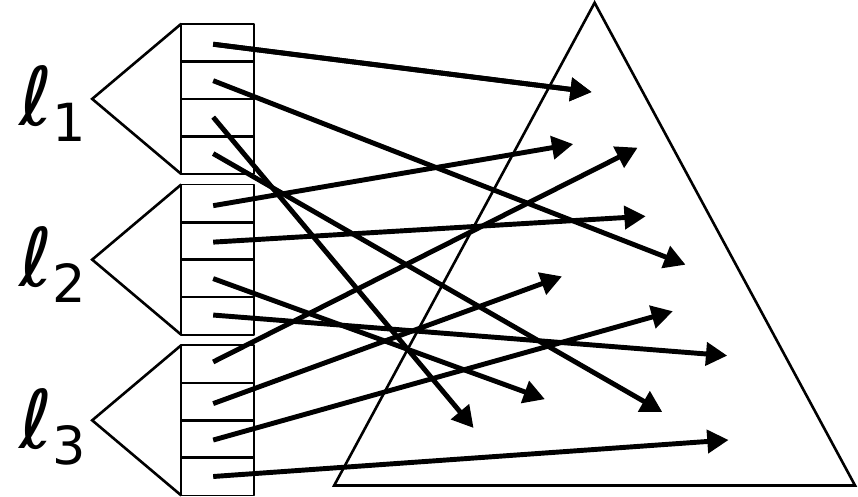}
  \caption{Indexing the AST by Label}
  \label{fig:labelIndex}
  \trimfigurespacing
\end{figure}

\subsection{Bolting DBToaster onto a Compiler}
DBToaster~\cite{DBLP:journals/vldb/KochAKNNLS14} in particular is designed for embedded use.
It compiles a set of queries down to a C++ or Scala data structure that maintains the query results.
The data structure exposes insert, delete, and update operations for each source relation; and materializes the results of each query into an iterable collection.
One strategy for improving compiler performance is to make the minimum changes required (i.e., ``bolt-on'') to allow the compiler to use an incremental view maintenance data structure generated by DBToaster:

\begin{enumerate}
  \item The reduction above generates SQL queries for each pattern-match query used by the optimizer.
  \item DBToaster builds a view maintenance data structure.
  \item The compiler is instrumented to register changes in the AST with the view maintenance data structure.
  \item Iterative searches in the optimizer for candidate AST nodes are replaced with a constant-time lookup on the view maintenance data structure.
\end{enumerate}

As we show in \Cref{sec:evaluation}, this approach significantly outperforms naive iterative AST scans.  
Although DBToaster requires maintaining supplemental data structures, the overhead of maintaining these structures is negligible compared to the benefit of constant-time pattern match results.

Nevertheless, there are three major shortcomings to this approach.
First, DBToaster effectively maintains a shadow copy of the entire AST --- at least the subset that affects pattern-matching results.
Second, DBToaster aggressively caches intermediate results.
\revised{
For example, our running example requires materializing 2 additional view queries, and this number grows combinatorially with the join width.  
}
Finally, DBToaster-generated view structures register updates at the granularity of individual node insertions/deletions, making it impossible for them to take advantage of the fact that most rewrites follow very structured patterns.
For a relatively small number of pattern-matches, the memory use of the compiler with a DBToaster view structure bolted on, increases by a factor of 2.5$\times$.
Given that memory consumption is already a pain point for large ASTs, this is not viable.

Before addressing these pain points, we first assess why they arise.
First, DBToaster-generated view maintenance data structures are self-contained.  
When an insert is registered, the structure needs to preserve state for later use.  
Although unnecessary fields are projected away, this still amounts to a shadow copy of the AST.
Second, DBToster has a heavy focus on aggregate queries.
Caching intermediate state allows aggressive use of aggregation and selection push-down into intermediate results, both reducing the amount of state maintained and the work needed to maintain views.

Both benefits are of limited use in pattern-matching on ASTs.
Pattern matches are SPJ queries, mitigating the value of aggregate push-down.
The value of selection push-down is mitigated by the AST's implicit foreign key constraints: each child has a single parent and each child attribute references at most one child.
Unlike a typical join where a single record may join with many results, here a single node only participates in a single join result\footnote{To clarify, a node may participate in multiple join results in different positions in the pattern match, but only in one result at the same position.}.
This also limits the value of materializing intermediate state for the sake of cache locality when updating downstream relations.

In summary, for ASTs, the cached state is either redundant or minimally beneficial.
Thus a view maintenance scheme designed specifically for compilers should be able to achieve the same benefits, but without the memory overhead.



\section{Pattern Matching on a Space Budget}
\label{sec:system}

We have a set of patterns $\Pattern_1, \ldots, \Pattern_m$ and an evolving abstract syntax tree $\Node$.  
Our goal is, given some $\Pattern_k$, to be able to obtain a single, arbitrary element of the set $\Pattern_k(\Node)$ as quickly as possible.
Furthermore, this should be possible without significant overhead as $\Node$ evolves into $\Node'$, $\Node''$, and so forth.
Recall that there are three properties that have to hold for a node $\Node$ to match $\Pattern$: 
(i) The node and pattern labels must match, 
(ii) Any recursively nested patterns must match, and
(iii) The constraint must hold over the node and its descendants.

\subsection{Indexing Labels}
\label{sec:labelIndexing}

A standard first step to query optimization is indexing, for example with a secondary index on the node labels as illustrated in \Cref{fig:labelIndex}.
For each node label, the index maintains pointers to all nodes with that label.
Updates to the AST are propagated into the index.
Pattern match queries can use this index to scan a subset of the AST that includes only nodes with the appropriate label, as shown in \Cref{alg:indexLookup}.


\begin{algorithm}
 \KwInput{$\Node \in \NodeDomain$, $\Pattern \in \PatternDomain$, $\texttt{Index}_\Node : \Label \rightarrow \setof{\Descendants{\Node}}$}
 \KwOutput{$\Node_{match} \in \Descendants{\Node}$}
 \If{$\Pattern = \AnyNode$}{
    \Return{$\Node_{match}\leftarrow \Node$}
 }
 \ElseIf{$\Pattern = \PatternNode{\Label}{\NodeVariable}{[\Pattern_1, \ldots, \Pattern_n]}{\Constraint}$}{
    \For{$\Node_{idx} \in \texttt{Index}_\Node[\Label]$}{
      \If{$\Pattern(\Node) = \True, \Scopefunction$}{
        \Return{$\Node_{match} \leftarrow \Node_{idx}$}
      }
    }
 }
\caption{\texttt{IndexLookup}($\Node, \Pattern, \texttt{Index}_{\Node}$)}
\label{alg:indexLookup}
\end{algorithm}

Indexing the AST by node label is simple and has a relatively small memory overhead: approximately 28 bytes per AST node using the C++ standard library \lstinline{unordered_set}.
Similarly, the maintenance overhead is low --- one hash table insert and/or remove per AST node changed.


\begin{example}
\revised{
To find matches for the rule of \Cref{ex:inlineArithmetic}, we retrieve a list of all \texttt{Arith} nodes from the index and iteratively check each for a pattern match.
Note that this approach only supports filtering on labels; Recursive matches and constraints both need to be re-checked with each iteration.
}
\end{example}


\subsection{Incremental View Maintenance}
\revised{
While indexing works well for single-node patterns, recursive patterns require a heavier-weight approach.
Concretely, when a node in the AST is updated, we need to figure out which new pattern matches the update creates, and which pattern matches it removes.  
As we saw in \Cref{sec:bolton}, this could be accomplished by ``joining'' the updated node with all of the other nodes that could participate in the pattern.
However, to compute these joins efficiently, DBToaster and similar systems need to maintain a significant amount of supporting state: 
(i) The view itself,
(ii) Intermediate state needed to evaluate subqueries efficiently
(iii) A shadow copy of the AST.
The insight behind \systemname is that the latter two sources of state are unnecessary when the AST is already available:
(i) Subqueries (inter-node joins) reduce to pointer chasing when the AST is available, and 
(ii) A shadow copy of the AST is unnecessary if the IVM system can navigate the AST directly.
}


\revised{
We begin to outline \systemname in \Cref{sec:IVM} by defining IVM for immutable (functional) ASTs.
This simplified form of the IVM problem has a useful property:
When a node is updated, all of its ancestors are updated as well.
Thus, we are guaranteed that the root of a pattern match will be part of the change set (i.e., $\Delta D$) and can restrict our search accordingly.
We then refine the approach to mutable ASTs, where only a subset of the tree is updated.
Then, in \Cref{sec:inlining} we describe how declarative specifications of rewrite rules can be used to streamline the derivation of update sets and to eliminate unnecessary checks.
Throughout \Cref{sec:IVM,sec:inlining}, we focus on the case of a single pattern query, but note that this approach generalizes trivially to multiple patterns.
}


\section{AST-Optimized IVM}
\label{sec:IVM}

\newcommand{\DomainOf}[1]{\textbf{dom}(#1)}
\newcommand{\GenericMultiset}{\mathbf M}

We first review a generalization of multisets proposed by Blizard~\cite{DBLP:journals/ndjfl/Blizard90} that allows for elements with negative multiplicities.
A generalized multiset $\GenericMultiset : \DomainOf{\GenericMultiset} \rightarrow \mathbb Z$ maps a set of elements from a domain $\DomainOf{\GenericMultiset}$ to an integer-valued multiplicity.
We assume finite-support for all generalized multisets: only a finite number of elements are mapped to non-zero multiplicities.
Union on a generalized multiset, denoted $\oplus$, is defined by summing multiplicities.  
$$(\GenericMultiset_1 \oplus \GenericMultiset_2)(x) \isDefinedAs \GenericMultiset_1(x) + \GenericMultiset_2(x)$$
Difference, denoted $\ominus$, is defined analogously:
$$(\GenericMultiset_1 \ominus \GenericMultiset_2)(x) \isDefinedAs \GenericMultiset_1(x) - \GenericMultiset_2(x)$$
We write $x \in \GenericMultiset$ as a shorthand for $\GenericMultiset(x) \neq 0$.
When combining sets and generalized multisets, we will abuse notation and lift sets to the corresponding generalized multiset, where each of the set's elements is mapped to the integer 1.



A view $\View{\Pattern}$ is a generalized multiset.
We define the correctness of a view relative to the root of an AST.
Without loss of generality, we assume that any node appears at most once in any AST.

\begin{definition}[View Correctness]
A view $\View{\Pattern}$ is correct for an AST node $\Node$ if the view is the generalized multiset that maps the subset of $\Descendants{\Node'}$ that matches $\Pattern$ to $1$, and all other elements to $0$:
$$\View{\Pattern} = \gmcomprehension{\Node' \rightarrow 1}{\Node' \in \Pattern(\Node)}$$
\end{definition}
    
If we start with a view $\View{\Pattern}$ that is correct for the root of an AST $\Node$ and rewrite the AST's root to $\Node'$, the view should update accordingly.
We initially assume that we have available a delta between the two ASTs (i.e., the difference $\Descendants{\Node'} \ominus \Descendants{\Node}$).
This delta is generally small for a rewrite, including only the nodes of the rewritten subtree and their ancestors.
We revisit this assumption in the following section.
\Cref{Algo:viewmaintenance} shows a simple algorithm for maintaining the $\View{\Pattern}$, given a small change $\Delta$, expressed as a generalized multiset.

\begin{algorithm}
 \KwInput{$\Pattern \in \PatternDomain$, $\View{\Pattern} \in \gmsetof{\NodeDomain}$, $\Delta \in \gmsetof{\NodeDomain}$}
 \KwOutput{$\View{\Pattern}'$}
 $\View{\Pattern}' \leftarrow \View{\Pattern}$\\
 \For{$\Node_i \in \Delta$\hspace{2mm} \tcc{nodes with multiplicity $\neq 0$}} 
 {
     \If{$\Pattern(\Node_i) =  \True,\Scopefunction$}
     {
        $\View{\Pattern}' \leftarrow \View{\Pattern}' \oplus \gmsetof{\Node_i \rightarrow \Delta(\Node_i)}$
     }
 }
\caption{$\IVM(\Pattern, \View{\Pattern}, \Delta)$}
\label{Algo:viewmaintenance}
\end{algorithm}

\begin{example}
\label{ex:replacement}
\revised{
Consider the AST of \Cref{fig:exampleAST}, which contains five nodes, and our ongoing example rule.  
Let us assume that the left subtree is replaced by $\texttt{Const}(0)$ (e.g., if $\texttt{Var(y)}$ is resolved to $0$).  
The multiset of the corresponding delta is: 
{\small
\begin{multline*}
\hspace*{-3mm}\left\{\hspace{-0.5mm}\left|\;
  (\texttt{Const}, \{\texttt{val} \mapsto 0\}, []) \mapsto 1,\;\;\;
  (\texttt{Arith}, \{\texttt{op} \mapsto +\}, [\ldots]) \mapsto 1,
\right.\right.\\
  (\texttt{Const}, \{\texttt{val} \mapsto 2\}, []) \mapsto -1,\;\;\;
  (\texttt{Var}, \{\texttt{name} \mapsto \texttt{y}\}, []) \mapsto -1,
\\\left.\left.
  (\texttt{Arith}, \{\texttt{op} \mapsto \times\}, [\ldots]) \mapsto -1
\;\right|\hspace{-0.5mm}\right\}
\end{multline*}
}
Only one of the nodes with nonzero multiplicities matches, making the update:
$\gmsetof{
(\texttt{Arith}, \{\texttt{op} \mapsto +\}, [\ldots]) \mapsto 1
}$
}
\end{example}

For \Cref{Algo:viewmaintenance} to be correct, we need to show that it computes exactly the update to $\View{\Pattern}$.

\begin{lemma}[Correctness of IVM]
\label{thm:ivmIsCorrect}
Given two ASTs $\Node$ and $\Node'$ and assuming that $\View{\Pattern}$ is correct for $\Node$, 
then the generalized multiset returned by  
$\IVM(\Pattern, \View{\Pattern}, \Descendants{\Node'} \ominus \Descendants{\Node})$ 
is correct for $\Node'$.
\end{lemma}
\IfTechReport{
   \begin{proof}
   Lets denote the generalized multiset returned from $\IVM(\Pattern, \View{\Pattern}, \Descendants{\Node'} \ominus \Descendants{\Node})$ as $\View{\Pattern}^{'}$.
  To prove that $\View{\Pattern^{'}}$ is correct we examine the multiplicity of a arbitrary node $\Node''$.\\
  (i) If $\Node'' \in \Descendants{\Node}$, $\Node'' \notin \Descendants{\Node'}$ and $\Pattern,\Node''\mapsto \True,\Scopefunction$
  {\footnotesize
  \begin{align*}
  \Descendants{N'}\ominus\Descendants{N}(\Node'') &=& -1\\
  \View{\Pattern}(\Node'') &=& 1 \\
  \View{\Pattern}' &=& \View{\Pattern}(\Node'') \oplus(\Descendants{N'}\ominus\Descendants{N})(\Node'')\\
     &=&0\\
     &=&\Pattern(\Node')
  \end{align*}
  }
  (ii) If $\Node'' \in \Descendants{\Node}$, $\Node'' \notin \Descendants{\Node'}$ and $\Pattern,\Node''\mapsto \False,\Scopefunction$
  {\footnotesize
  \begin{align*}
  \Descendants{N'}\ominus\Descendants{N}(\Node'') &=& -1 \\
  \View{\Pattern}(\Node'') &=& 0 \\
  \View{\Pattern}' &=& \View{\Pattern}(\Node'') \\
     &=&0\\
     &=&\Pattern(\Node')
  \end{align*}
  }
  (iii) If $\Node'' \notin \Descendants{\Node}$, $\Node'' \in \Descendants{\Node'}$ and $\Pattern,\Node''\mapsto \True,\Scopefunction$
  {\footnotesize
  \begin{align*}
  \Descendants{N'}\ominus\Descendants{N}(\Node'') &=& 1\\
  \View{\Pattern}(\Node'') &=& 0 \\
  \View{\Pattern}' &=& \View{\Pattern}(\Node'') \oplus(\Descendants{N'}\ominus\Descendants{N})(\Node'')\\
     &=&1\\
     &=&\Pattern(\Node')
  \end{align*}
  }
  (iv)If $\Node'' \notin \Descendants{\Node}$, $\Node'' \in \Descendants{\Node'}$ and $\Pattern,\Node''\mapsto \False,\Scopefunction$
  {\footnotesize
  \begin{align*}
  \Descendants{N'}\ominus\Descendants{N}(\Node'') &=& 1\\
  \View{\Pattern}(\Node'') &=& 0 \\
  \View{\Pattern}' &=& \View{\Pattern}(\Node'')\\
     &=&0\\
     &=&\Pattern(\Node')
  \end{align*}
  }
  (v) If $\Node'' \in \Descendants{\Node}$, $\Node'' \in \Descendants{\Node'}$ and $\Pattern,\Node''\mapsto \True,\Scopefunction$
  {\footnotesize
  \begin{align*}
  \Descendants{N'}\ominus\Descendants{N}(\Node'') &=& 0\\
  \View{\Pattern}(\Node'') &=& 1 \\
  \View{\Pattern}' &=& \View{\Pattern}(\Node'') \oplus(\Descendants{N'}\ominus\Descendants{N})(\Node'')\\
     &=&1\\
     &=&\Pattern(\Node')
  \end{align*}
  }
  (vi) If $\Node'' \in \Descendants{\Node}$, $\Node'' \in \Descendants{\Node'}$ and $\Pattern,\Node''\mapsto \False,\Scopefunction$
  {\footnotesize
  \begin{align*}
  \Descendants{N'}\ominus\Descendants{N}(\Node'') &=& 0\\
  \View{\Pattern}(\Node'') &=& 0 \\
  \View{\Pattern}' &=& \View{\Pattern}(\Node'')\\
     &=&0\\
     &=&\Pattern(\Node')
  \end{align*}
  }
  (vii )If $\Node'' \notin \Descendants{\Node}$, $\Node'' \notin \Descendants{\Node'}$ and $\Pattern,\Node''\mapsto \True,\Scopefunction$
  {\footnotesize
  \begin{align*}
  \Descendants{N'}\ominus\Descendants{N}(\Node'') &=& 0\\
  \View{\Pattern}(\Node'') &=& 0 \\
  \View{\Pattern}' &=& \View{\Pattern}(\Node'') \oplus(\Descendants{N'}\ominus\Descendants{N})(\Node'')\\
     &=&0\\
     &=&\Pattern(\Node')
  \end{align*}
  }
  (viii) If $\Node'' \notin \Descendants{\Node}$, $\Node'' \notin \Descendants{\Node'}$ and $\Pattern,\Node''\mapsto \False,\Scopefunction$
  {\footnotesize
  \begin{align*}
  \Descendants{N'}\ominus\Descendants{N}(\Node'') &=& 0\\
  \View{\Pattern}(\Node'') &=& 0 \\
  \View{\Pattern}' &=& \View{\Pattern}(\Node'') \\
     &=&0\\
     &=&\Pattern(\Node')
  \end{align*}
  }
  Since for all 8 possibilities \Cref{Algo:viewmaintenance} computes correctly the multiplicity of node $\Node''$ in the resultant of $\IVM(\Pattern, \View{\Pattern}, \\ \Descendants{\Node'} \ominus \Descendants{\Node}$ as it would be in $\Descendants{\Node'}$ \Cref{Algo:viewmaintenance} is correct.
  \end{proof} 
}{
  \begin{sketch}
We examine the multiplicity of each node $\Node''$ over different cases of $\Node'' \in \Descendants{\Node}, \Node'' \in \Descendants{\Node'}$ and $\Pattern(\Node'') = \True/\False,\Scopefunction$ and show equivalence between the incrementally maintained and naively recomputed view.
If $\Pattern(\Node'') = \False,\Scopefunction$ then the multiplicity remains unchanged for node $\Node''$. Otherwise, The condition on line 2 would apply and the multiplicity gets recalculated according to line 3. 
  \end{sketch}
  The full proof appears in an associated tech report ~\cite{balakrishnan2021treetoaster}.
}

\subsection{Mutable Abstract Syntax Trees}
Although correct, \IVM{} assumes that the AST is immutable: When a node changes, each of its ancestors must be updated to reference the new node as well.
\revised{
  Even when \systemname is built into a compiler with immutable ASTs, many of these pattern matches will be redundant.
  By lifting this restriction (if in spirit only), we can decrease the overhead of view maintenance by reducing the number of nodes that need to be checked with each AST update.
}
To begin, we create a notational distinction between the root node $\ASTroot$ and the node being replaced $\Patternroot$.
For clarity of presentation, we again assume that any node $\Patternroot$ occurs at most once in $\ASTroot$.
$\ASTroot[\Patternroot \backslash \Patternroot']$ is the node resulting from a replacement of $\Patternroot$ with $\Patternroot'$ in $\ASTroot$:
$$\ASTroot[\Patternroot \backslash \Patternroot'] = \begin{cases}
\Patternroot' \hspace{15mm} & \textbf{if } \ASTroot = \Patternroot\\
\multicolumn{2}{l}{\hspace*{-1.5mm}
\NodeExplicit{\Label}{\AnnotationMap}{[\Node_1[\Patternroot \backslash \Patternroot'], \ldots, \Node_n[\Patternroot \backslash \Patternroot']]}
} \\
& \textbf{if } \ASTroot = \NodeExplicit{\Label}{\AnnotationMap}{[\Node_1, \ldots, \Node_n]}
\end{cases}$$
We also lift this notation to collections:
$$\View{}[\Patternroot \backslash \Patternroot'] = \gmcomprehension{\Node[\Patternroot \backslash \Patternroot'] \rightarrow c}{(\Node \rightarrow c) \in \View{}}$$
We emphasize that although this notation modifies each node individually, this complexity appears only in the analysis.
The underlying effect being modeled is a single pointer swap.

\begin{example}
\revised{
The replacement of \Cref{ex:replacement} is written: 
$$\ASTroot[\;{\small (\texttt{Arith}, \{\texttt{op} \mapsto \times\},[\ldots])} \; \backslash \; {\small (\texttt{Const}, \{\texttt{val} \mapsto 0\}, [])}\;]$$
In the mutable model, the root node itself does not change.
}
\end{example}

\begin{definition}[Pattern Depth]
The depth $\Patterndepth{\Pattern}$ of a pattern $\Pattern$ is the number of edges along the longest downward path from root of the pattern to an arbitrary pattern node $\Pattern_i$.
$$\Patterndepth{\Pattern} = \begin{cases}
     0 &\textbf{if } \Pattern = \AnyNode\\
    1+\underset{i \in [n]}{max}(\Patterndepth{q_i}) &\textbf{if } \Pattern = \PatternNode{\Label}{\NodeVariable}{[\Pattern_1, \ldots, \Pattern_n]}{\Constraint}
\end{cases}$$
\end{definition}

The challenge posed by mutable ASTs is that the modified node may make one of its ancestors eligible for a pattern-match.  
However, as we will show, only a bounded number of ancestors are required.  
Denote by $\Ancestor{i}{\Node}$ the $i$th ancestor of $\Node$\footnote{We note that ASTs do not generally include ancestor pointers.  The ancestor may be derived by maintaining a map of node parents, or by extending the AST definition with parent pointers.}.
The maximal search set, which we now define, includes all nodes that need to be checked for matches.

\begin{definition}[Maximal Search Set]
Let $\Patternroot$ and $\Patternroot'$ be an arbitrary node in the AST and its replacement.
The maximal search set for $\Patternroot$ and $\Patternroot'$ and pattern $\Pattern$, $\ceil{\Patternroot, \Patternroot'}_\Pattern$ is the difference between the generalized multiset of the respective nodes, their descendents, and their ancestors up to a height of $\Patterndepth{\Pattern}$.
{\small
\begin{multline*}
\ceil{\Patternroot, \Patternroot'}_\Pattern \isDefinedAs 
\Descendants{\Patternroot} 
\oplus \gmcomprehension{\Ancestor{i}{\Patternroot} \rightarrow 1}{i \in [n]}\\
\ominus \Descendants{\Patternroot'} 
\ominus \gmcomprehension{\Ancestor{i}{\Patternroot'} \rightarrow 1}{i \in [n]}
\end{multline*}
}
\end{definition}

\begin{lemma}
Let $\ASTroot$ be the root of an AST, $\Pattern$ be a pattern, and $\Patternroot$ and $\Patternroot'$ be an arbitrary node in the AST and its replacement.  If $\View{\Pattern}$ is correct for $\ASTroot$.
and $\View{\Pattern}' = \IVM(\Pattern, \View{\Pattern}, \ceil{\Patternroot, \Patternroot'}_\Pattern)$, then $\View{\Pattern}'[\Patternroot \backslash \Patternroot']$ is correct for $\ASTroot[\Patternroot \backslash \Patternroot']$
\end{lemma}
\IfTechReport{
  \begin{proof}
  Differs from the proof of \Cref{thm:ivmIsCorrect} in three additional cases.  
  If $\Pattern, \Node'' \mapsto \tuple{\False, \Scopefunction}$, then $\Pattern(\ASTroot) = \ASTroot[\Patternroot \backslash \Patternroot'] = \View{\Pattern} = 0$.  The condition on line 2 is false, and the multiplicity is unchanged at 0. 
  Otherwise, if $\Node'' \in \Descendants{\ASTroot}$ then $\Node''[\Patternroot \backslash \Patternroot'] \in \Descendants{\ASTroot[\Patternroot \backslash \Patternroot']}$ by definition.
  Here, there are two possibilities: Either \Node'' is within the $\Patterndepth{\Pattern}$-high ancestors of $\Patternroot$ or not.  
  In the former case, both $\Node''$ and $\Node''[\Patternroot \backslash \Patternroot']$ appear in $\ceil{\Patternroot, \Patternroot'}_\Pattern$ with multiplicities $1$ and $-1$ respectively, and the proof is identical to \Cref{thm:ivmIsCorrect}.  
  We prove the latter case by recursion, by showing that if $\Node''$ is not among the $\Patterndepth{\Pattern}$ immediate ancestors and $\Pattern,\Node'' \mapsto \tuple{x, \Gamma}$, then $\Pattern,\Node''[\Patternroot \backslash \Patternroot']$ does as well.
  .  The base case is a pattern depth of 0, or $\Pattern = \AnyNode$.  
  This pattern always evaluates to $\tuple{\True, \emptyset}$ regardless of input, so the condition is satisfied.
  For the recursive case, we assume that the property holds for any $\Pattern$ and $\Node'''$ not among the $d-1$th ancestors of $\Patternroot$.  
  Since $d>1$, $\Patternroot \neq \Node''$ and the precondition for Pattern Rule 2.1 is guaranteed to be unchanged.
  If a pattern has depth $d$, none of its children have depth more than $d-1$ so we have for each of the pattern's children that if $\Pattern_i,\Node_i \mapsto \tuple{x_i, \Gamma_i}$ then $\Pattern_i,\Node_i[\Patternroot \backslash \Patternroot'] \mapsto \tuple{x_i, \Gamma_i}$, and the preconditions for Pattern Rules 2.1 and 2.3 are unchanged.  
  Likewise, since both inputs map to identical gammas and $\Patternroot \neq \Node''$, the preconditions for Pattern rule 2.4 are unchanged.
  Since the preconditions for all relevant pattern-matching rules are unchanged, the condition holds at a depth of $d$.
  \end{proof}
}{
  \begin{sketch}
  Shown by building on the proof of \Cref{thm:ivmIsCorrect} with a recursive proof that changing a subtree can not affect the matchability of a node more than $\Patterndepth{\Pattern}$ ancestors above. The full proof appears in an associated tech report ~\cite{balakrishnan2021treetoaster}.
  \end{sketch}
}

\begin{example}
The pattern depth of our running example is 1.  Continuing the prior example, only the node, its 1-ancestor (i.e., parent), and the 1-descendents (i.e., children) of the replacement node would need to be examined for view updates.
\end{example}

\section{Inlining into Rewrite Rules}
\label{sec:inlining}

\Cref{Algo:viewmaintenance} takes the set of changed nodes as an input.
In principle, this information could be obtained by manually instrumenting the compiler to record node insertions, updates, and deletions.
However, many rewrite rules are structured: The rule replaces exactly the matched set of nodes with a new subtree.  
Unmodified descendants are re-used as-is, and with mutable ASTs a subset of the ancestors of the modified node are re-used as well.  
\systemname provides a declarative language for specifying the output of rewrite rules.
This language serves two purposes.
In addition to making it easier to instrument node changes for \systemname, declaratively specifying updates opens up several opportunities for inlining-style optimizations to the view maintenance system.
The declarative node generator grammar follows:
\begin{mathpar}
\GenerateDomain : \GenerateNode{\LabelDomain}{\;\overline{\AtomDomain}}{\;\overline \GenerateDomain} 
            \gOR \GenerateExisting{\NodeVariableDomain}
\end{mathpar}
A \GenerateNodeName term indicates the creation of a new node with the specified label, attributes, and children.
Attribute values are populated according to a provided attribute scope $\Scopefunction : \NodeVariableDomain \rightarrow \AnnotationVariableDomain \rightarrow \ValueDomain$.
A \GenerateExistingName term indicates the re-use of a subtree from the previous AST, provided by a node scope $\mu : \NodeVariableDomain \rightarrow \NodeDomain$.
Node generators are evaluated by the $\DoGenerate{\cdot}_{\Scopefunction, \mu} : \GenerateDomain \rightarrow \NodeDomain$ operator, defined as follows:
{\small
$$\DoGenerate{\Generate}_{\Scopefunction,\mu} = \begin{cases}
  \mu(\NodeVariable) \hspace{10mm} & \textbf{if } \Generate = \GenerateExisting{\NodeVariable}\\
  \multicolumn{2}{l}{\hspace*{-1.5mm}
    \NodeExplicit{\Label}{\{\Atom_1(\Scopefunction), \ldots, \Atom_k(\Scopefunction)\}}{\DoGenerate{\Generate_1}_{\Scopefunction,\mu}, \ldots, \DoGenerate{\Generate_n}_{\Scopefunction,\mu}} 
  }\\
    & \textbf{if } \Generate = \GenerateNode{\Label}{[\Atom_1, \ldots, \Atom_k]}{[\Generate_1, \ldots, \Generate_n]}
\end{cases}$$
}

A declaratively specified rewrite rule is given by a 2-tuple: $\tuple{\Pattern,\Generate} \in \PatternDomain \times \GenerateDomain$, a match pattern describing the nodes to be removed from the tree, and a corresponding generator describing the nodes to be inserted back into the tree as replacements.
As a simplification for clarity of presentation, we require that \GenerateExistingName nodes reference nodes matched by \AnyNode{} patterns.  
Define the set of matched node pairs as the set
{\small
\begin{multline*}
\MatchPairs{\Pattern}{\Patternroot} = \setof{\tuple{\Pattern, \Patternroot}}\cup \ldots\\ \ldots \begin{cases}
\setof{\tuple{\AnyNode, \Patternroot}} & \textbf{if } \Pattern = \AnyNode\\
\underset{k \in [n]}{\bigcup} \MatchPairs{\Pattern_k}{\Node_k} 
          & \textbf{if } \Pattern = \PatternNode{\Label}{\AnnotationVariable}{[\Pattern_1, \ldots, \Pattern_n]}{\Constraint} \\[-3mm]
          & \hspace{4mm} \Patternroot = \NodeExplicit{\Label}{\AnnotationMap}{[\Node_1, \ldots, \Node_n]}
\end{cases}
\end{multline*}
}
A set of generated node pairs $\GeneratePairs{\Generate, \Scopefunction, \mu}$ is defined analogously relative to the node $\DoGenerate{\Generate}_{\Scopefunction,\mu}$

\begin{definition}[Safe Generators]
Let $\ASTroot$ be an AST root, $\Pattern$ be a pattern query, and $\Patternroot \in \Pattern(\ASTroot)$ be a node of the AST matching the pattern.
We call a generator $\Generate \in \GenerateDomain$ \emph{safe} for $\tuple{\Pattern,\Patternroot}$ iff $\Generate$ reuses exactly the wildcard matches of $\Pattern$.  Formally:
{\small
$$\tuple{\AnyNode, \Node} \in \MatchPairs{\Pattern}{\Patternroot} \Leftrightarrow
  \tuple{\GenerateExisting{\Node}, \Node} \in \GeneratePairs{\Generate, \Scopefunction, \mu}$$
}
\end{definition}

Let $\Generate \in \GenerateDomain$ be a generator that is safe for $\tuple{m, \Patternroot}$, where $m \in \PatternDomain$ is a pattern.  
The mutable update delta from $\ASTroot$ to $\ASTroot[\Patternroot \backslash \DoGenerate{\Generate}_{\Scopefunction, \mu}]$ is: 
\begin{multline*}
\Delta = \gmcomprehension{\Node' \rightarrow 1}{\tuple{\Generate', \Node'} \in \GeneratePairs{\Generate, \Scopefunction, \mu}} \ominus \\
\gmcomprehension{\Node' \rightarrow 1}{\tuple{\Pattern', \Node'} \in \MatchPairs{m}{\Patternroot}}
\end{multline*}
Note that the size of this delta is linear in the size of $\Generate$ and $m$.

\subsection{Inlining Optimizations}
Up to now, we have assumed that no information about the nodes in the update delta is available at compile time.
For declarative rewrite rules, we are no longer subject to this restriction.
The labels and structure of the nodes being removed and those being added are known at compile time.
This allows \systemname to generate more efficient code by eliminating impossible pattern matches.

\newcommand{\MatchGenerate}[2]{\MatchGenerateParent{#1}{#2}{0}}
\newcommand{\MatchGenerateParent}[3]{\texttt{Align}_{#3}(#1, #2)}

\begin{algorithm}
  \KwInput{$\Pattern \in \PatternDomain$, $\Generate \in \GenerateDomain$}
  \KwOutput{$f : \NodeDomain \mapsto \setof{\NodeDomain}$}
  \If{$\Pattern = \AnyNode \vee \Generate = \GenerateExisting{\mu}$}{
    $f' \leftarrow (\Node \mapsto \setof{\Node})$
  }
  \ElseIf{$\Pattern = \PatternNode{\Label}{\NodeVariable}{[\Pattern_1, \ldots, \Pattern_n]}{\Constraint} $} { 
    $\Generate = \GenerateNode{\Label'}{\NodeVariable'}{[\Generate_1, \ldots, \Generate_n]}$\;
    \If{$\MatchGenerate{\Pattern}{\Generate}$}{
      $f'' \leftarrow (\Node \mapsto \setof{\Node})$
    }
    \Else{
      $f'' \leftarrow (\Node \mapsto \emptyset)$
    }
    \For{$i \in [n]$}{
      $f_i \leftarrow \InlineGenerate(\Pattern, \Generate_i)$
    }
    $f' \leftarrow (\Node \mapsto f''(\Node) \cup \bigcup_{i \in [n]} f_i(\Node)$
  }
  $\mathcal{A} = \comprehension{i}{i \in [\Patterndepth{\Pattern}] \wedge \MatchGenerateParent{\Pattern}{\Generate}{i}}$\;
  $f \leftarrow (\Node \mapsto f'(\Node) \cup \comprehension{\Ancestor{i}{\Node}}{i \in \mathcal A}$
\caption{$\InlineGenerate(\Pattern, \Generate)$}
\label{Algo:inline}
\end{algorithm}

The process of elimination is outlined for generated nodes in \Cref{Algo:inline}.  
A virtually identical process is used for matching removed nodes.
The algorithm outputs a function that, given the generated replacement node (i.e., $\DoGenerate{\Generate}_{\Scopefunction, \mu}$) that is not available until runtime, returns the set of nodes that could match the provided pattern.
Matching only happens by label, as attribute values are also not available until runtime.
If the pattern matches anything or if the node is re-used (i.e., its label is not known until runtime), the node is a candidate for pattern match (Lines 1-2).
Otherwise, the algorithm proceeds in two stages.
It checks if a newly generated node can be the root of a pattern by recursively descending through the generator (Lines 3-11).
Finally, it checks if any of the node's ancestors (up to the depth of the pattern) could be the root of a pattern match by recursively descending through the pattern to see if the root of the generated node could match (Lines 12-13).
On lines 5 and 12, \Cref{Algo:inline} makes use of a recursive helper function: $\texttt{Align}$.  In the base case $\texttt{Align}_0$ checks if the input pattern and generator align -- whether they have equivalent labels at equivalent positions.
{\footnotesize
$$\MatchGenerateParent{\Pattern}{\Generate}{0} = \begin{cases}
\True  & \textbf{if } \Pattern = \AnyNode \vee \Generate = \GenerateExisting{\mu} \\
\False & \textbf{if } \Pattern = \PatternNode{\Label}{\AnnotationMap}{[\ldots]}{\Constraint}\\
       & \hspace{4mm}            \Generate = \GenerateNode{\Label'}{\NodeVariable}{[\ldots]} \wedge \Label \neq \Label'\\
\forall k : \MatchGenerateParent{\Pattern_k}{\Generate_k}{0}
       & \textbf{ if }\Pattern = \PatternNode{\Label}{\AnnotationMap}{[\ldots]}{\Constraint}\\
       & \hspace{4mm}            \Generate = \GenerateNode{\Label}{\NodeVariable}{[\ldots]}\\
\end{cases}$$
}
The recursive case $\texttt{Align}_{d}$ checks for the existence an alignment among the $d$th level descendants of the input pattern.
$$\MatchGenerateParent{\Pattern}{\Generate}{d} = 
\exists k : \MatchGenerateParent{\Pattern_k}{\Generate}{d-1}
$$

\begin{example}
Continuing the running example, only the $\texttt{Var}$ node appears in both the pattern and replacement.  
Thus, when a replacement is applied we need only check the parent of a replaced node for new view updates.
\end{example}

\section{Evaluation}
\label{sec:evaluation}


To evaluate \systemname, we built four IVM mechanisms into the \jitd~\cite{balakrishnan2019just,DBLP:conf/dbpl/BalakrishnanZK19} compiler, a JIT compiler for data structures built around a complex AST\footnote{The full result set of our runs is available at \url{https://github.com/UBOdin/jitd-synthesis/tree/master/treetoaster_scripts}}.
The \jitd compiler naturally works with large ASTs and requires low latencies, making it a compelling use case.
As such \jitd's provide an infrastructure to test \systemname.
Our tests compare: 
(i) The \jitd compiler's existing \textbf{Naive} iteration-based optimizer, 
(ii) \textbf{Index}ing labels, as proposed in \Cref{sec:labelIndexing}, 
(iii) \textbf{Classic}al incremental view maintenance implemented by bolting on a view maintenance data structure created by \dbt with the \texttt{--depth=1} flag, 
(iv) \textbf{DBT}oaster's full recursive view maintenance bolted onto the compiler, and
(v) \systemname (\textbf{TT})'s view maintenance built into the compiler.

Our experiments confirm the following claims:
(i) \systemname significantly outperforms \jitd's naive iteration-based optimizer,
(ii) \systemname matches or outperforms bolt-on IVM systems, while consiming significantly less memory,
(iii) On complex workloads, \systemname's view maintenance latency is half of bolt-on approaches,


\begin{figure}
\centering

\includegraphics[width=\columnwidth]{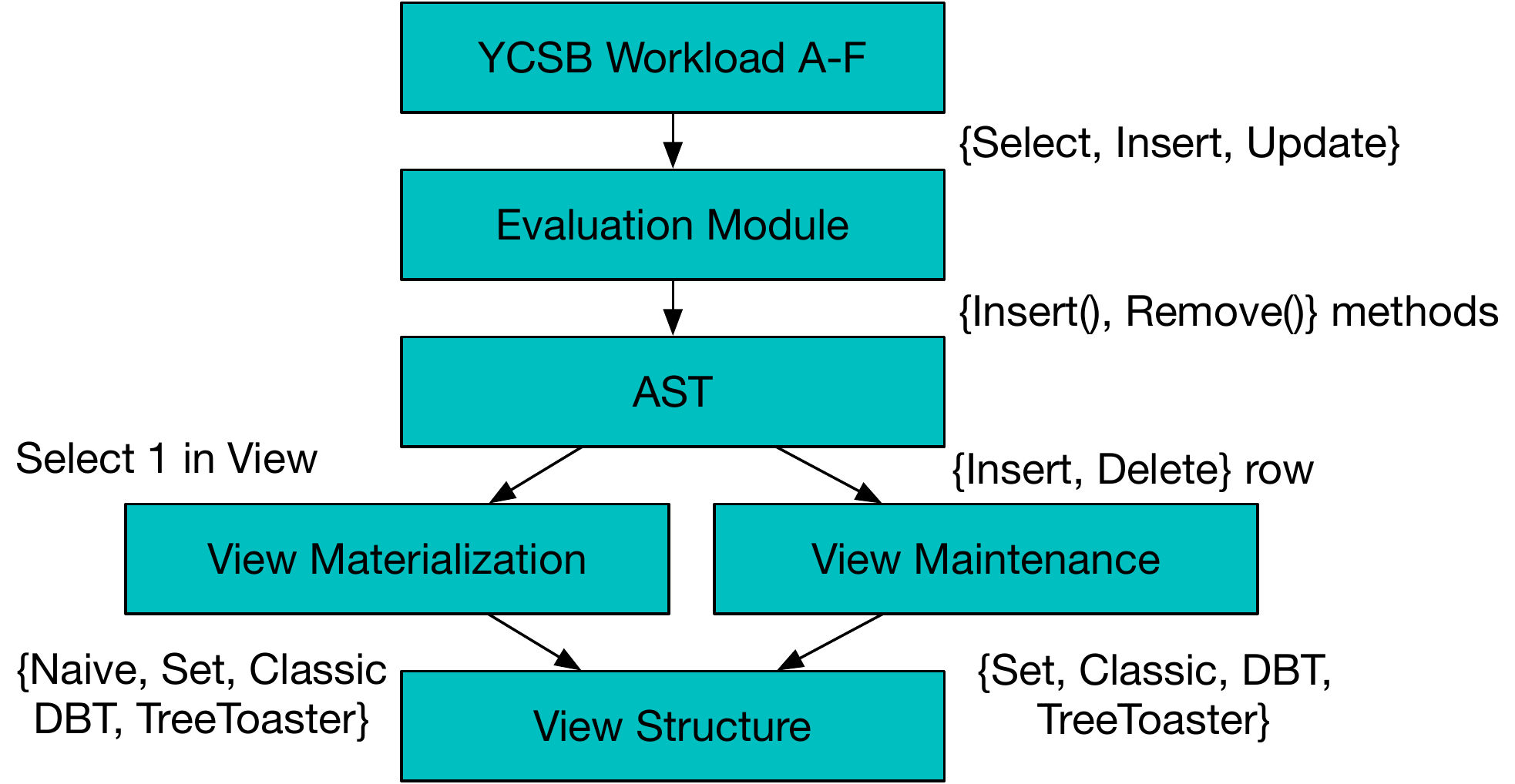}

\caption{Benchmark Infrastructure}
\label{fig:benchmark_system}
\trimfigurespacing
\end{figure}



\begin{figure*}
\centering
\begin{subfigure}{0.99\textwidth}
\includegraphics[trim={2cm 0 0 0},clip,width=0.99\textwidth]{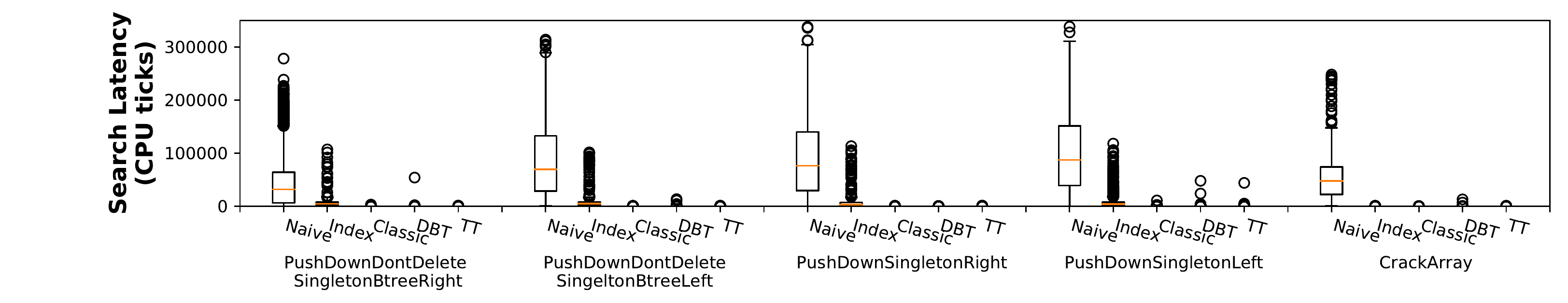}
\caption{Workload A}
\label{fig:search_performance_a}
\end{subfigure}
\begin{subfigure}{0.99\textwidth}
\includegraphics[trim={2cm 0 0 0},clip,width=0.99\textwidth]{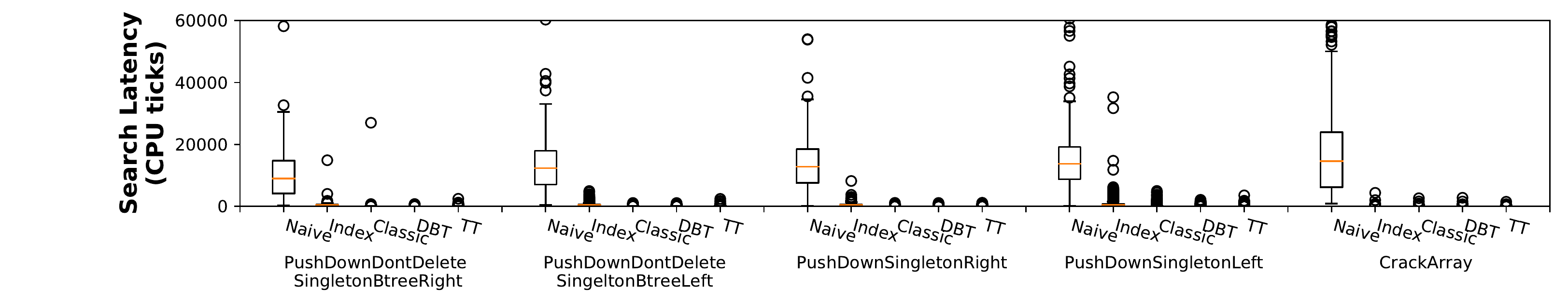}
\caption{Workload B}
\label{fig:search_performance_b}
\end{subfigure}
\begin{subfigure}{0.99\textwidth}
\includegraphics[trim={2cm 0 0 0},clip,width=0.99\textwidth]{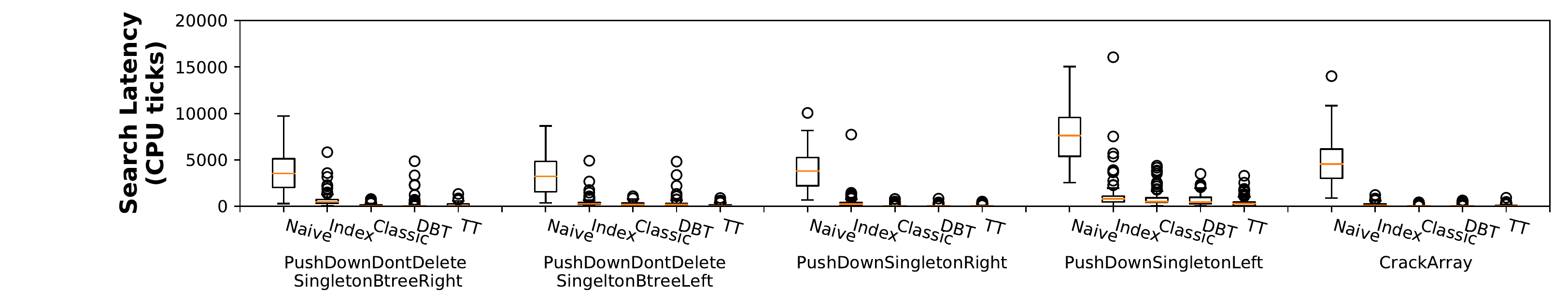}
\caption{Workload C}
\label{fig:search_performance_c}
\end{subfigure}
\begin{subfigure}{0.99\textwidth}
\includegraphics[trim={2cm 0 0 0},clip,width=0.99\textwidth]{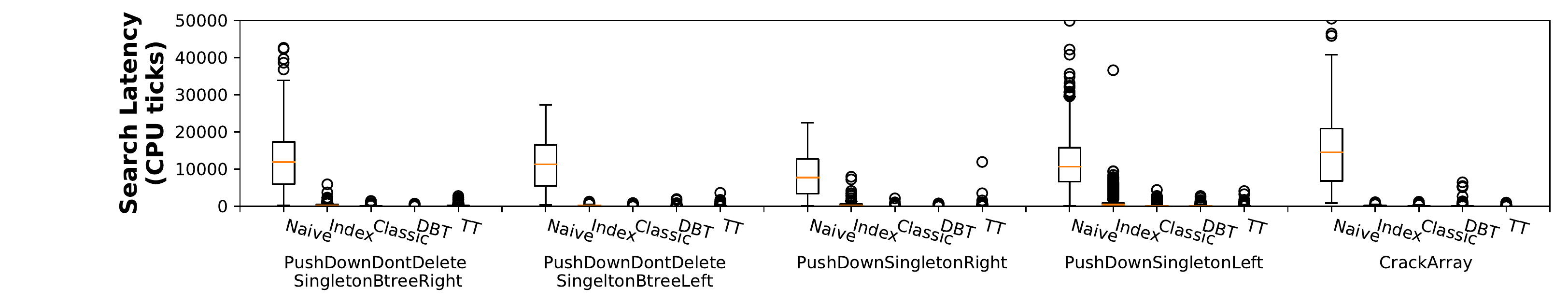}
\caption{Workload D}
\label{fig:search_performance_d}
\end{subfigure}
\begin{subfigure}{0.99\textwidth}
\includegraphics[trim={2cm 0 0 0},clip,width=0.99\textwidth]{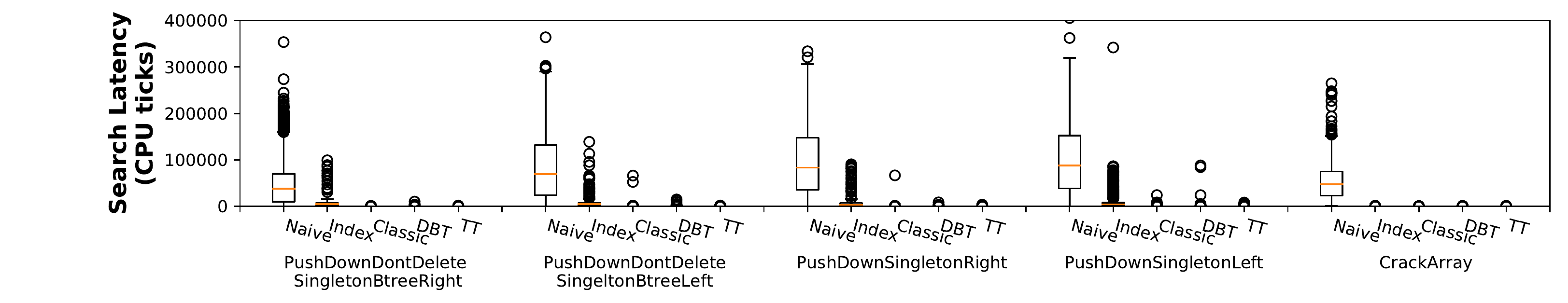}
\caption{Workload F}
\label{fig:search_performance_f}
\end{subfigure}
\caption{Relative Average Search Technique Performance by Rewrite Rule}
\label{fig:search_performance_all}
\end{figure*}




\newlength{\TrimFigureHeight}
\setlength{\TrimFigureHeight}{-1mm}
\begin{figure*}
\centering
\begin{subfigure}{0.99\textwidth}
\includegraphics[trim={2cm 0 0 0},clip,width=0.99\textwidth]{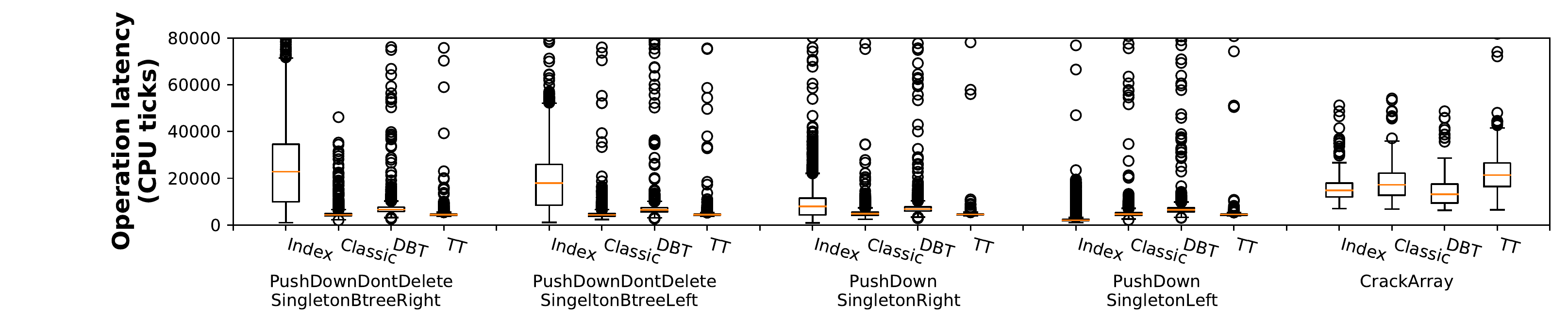}
\caption{Workload A}
\label{fig:combination_performance_a}
\end{subfigure}
\begin{subfigure}{0.99\textwidth}
\includegraphics[trim={2cm 0 0 0},clip,width=0.99\textwidth]{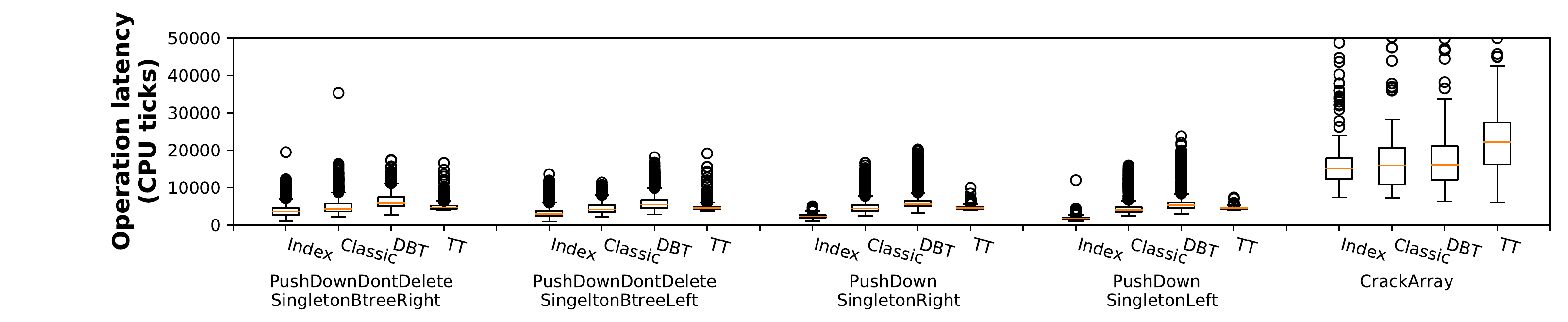}
\caption{Workload B}
\label{fig:combination_performance_b}
\end{subfigure}
\begin{subfigure}{0.99\textwidth}
\includegraphics[trim={2cm 0 0 0},clip,width=0.99\textwidth]{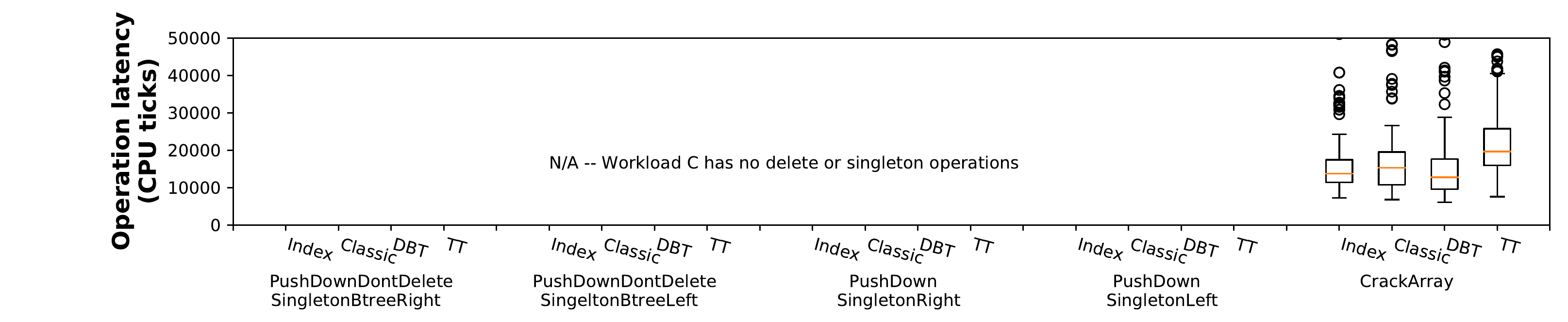}
\caption{Workload C}
\label{fig:combination_performance_c}
\end{subfigure}
\begin{subfigure}{0.99\textwidth}
\includegraphics[trim={2cm 0 0 0},clip,width=0.99\textwidth]{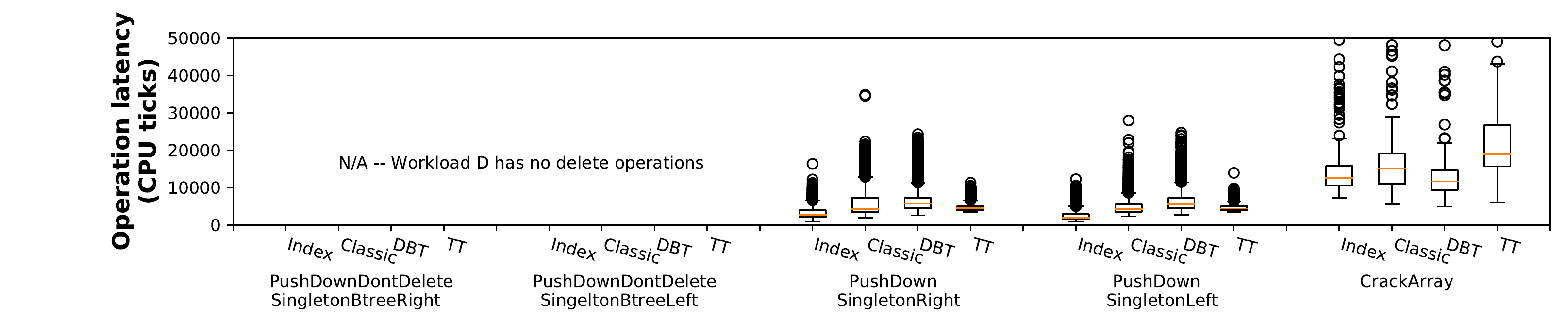}
\caption{Workload D}
\label{fig:combination_performance_d}
\end{subfigure}
\begin{subfigure}{0.99\textwidth}
\includegraphics[trim={2cm 0 0 0},clip,width=0.99\textwidth]{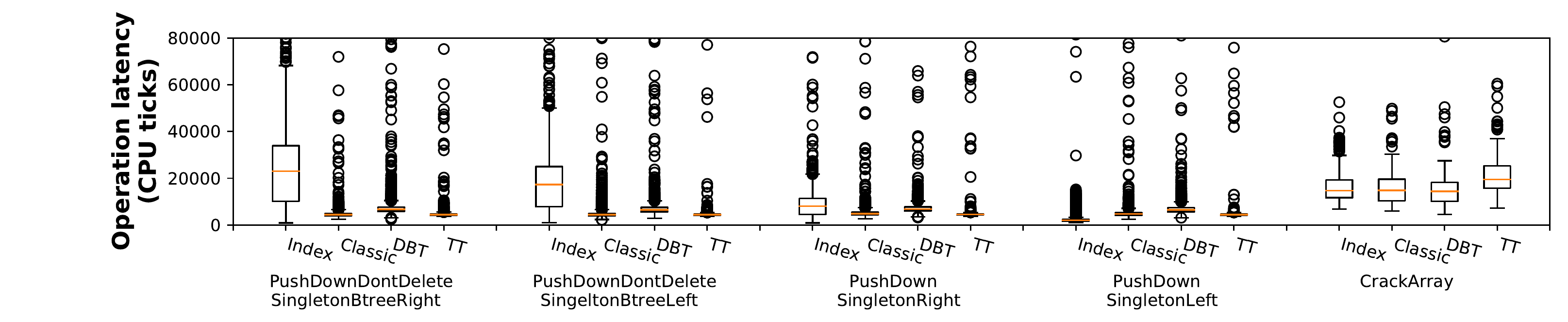}
\caption{Workload F}
\label{fig:combination_performance_f}
\end{subfigure}
\caption{Relative Total Search + Maintenance Cost by Rewrite Rule}
\label{fig:combination_performance_all}
\end{figure*}


\begin{figure}
\centering
\includegraphics[width=\columnwidth]{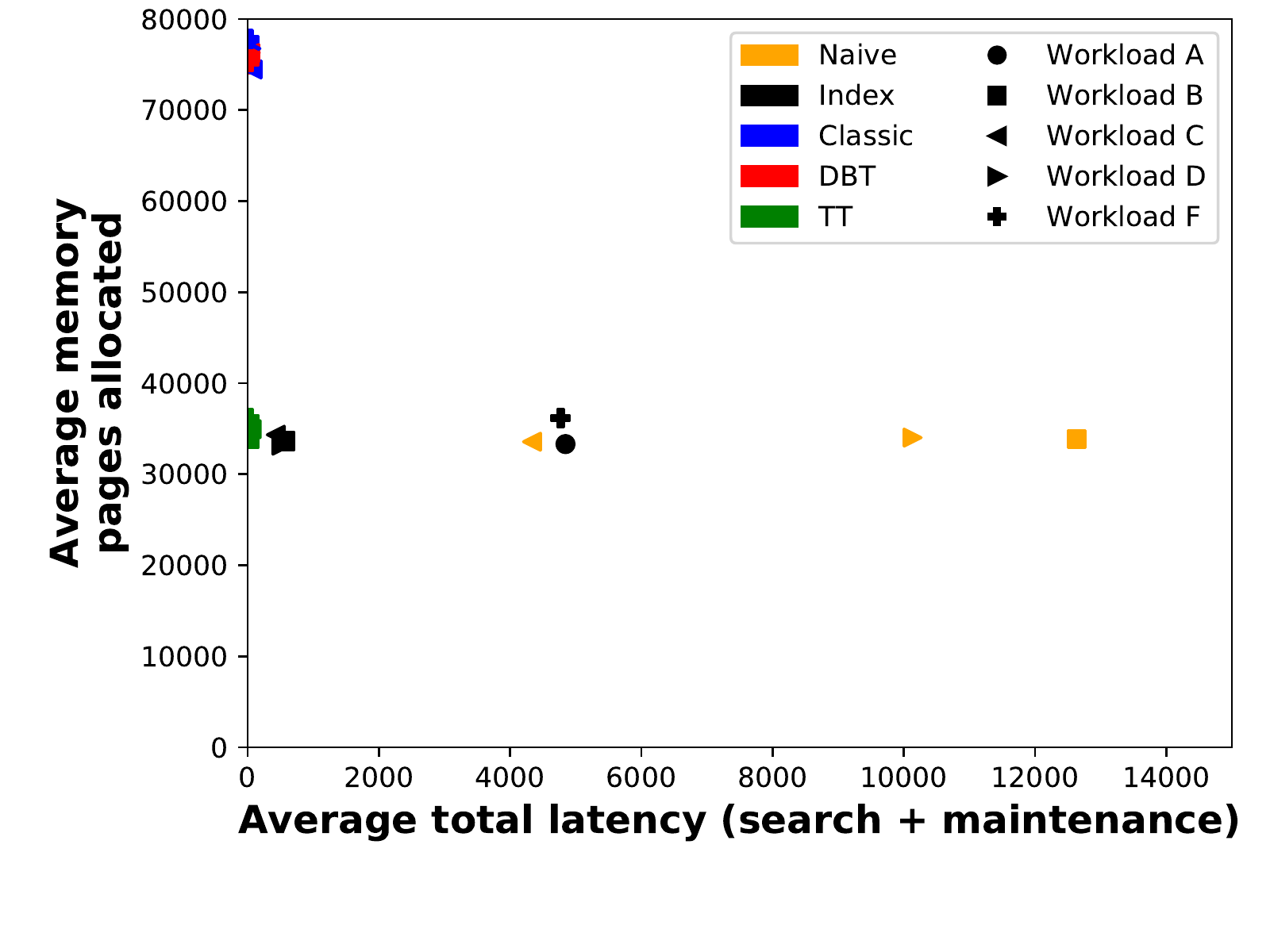}
\caption{Total Latency (search cost + maintenance operations) and Memory Use, by method and node type}
\label{fig:crossplot}
\trimfigurespacing
\end{figure}


\begin{figure}
\centering
\centering
\includegraphics[width=\columnwidth]{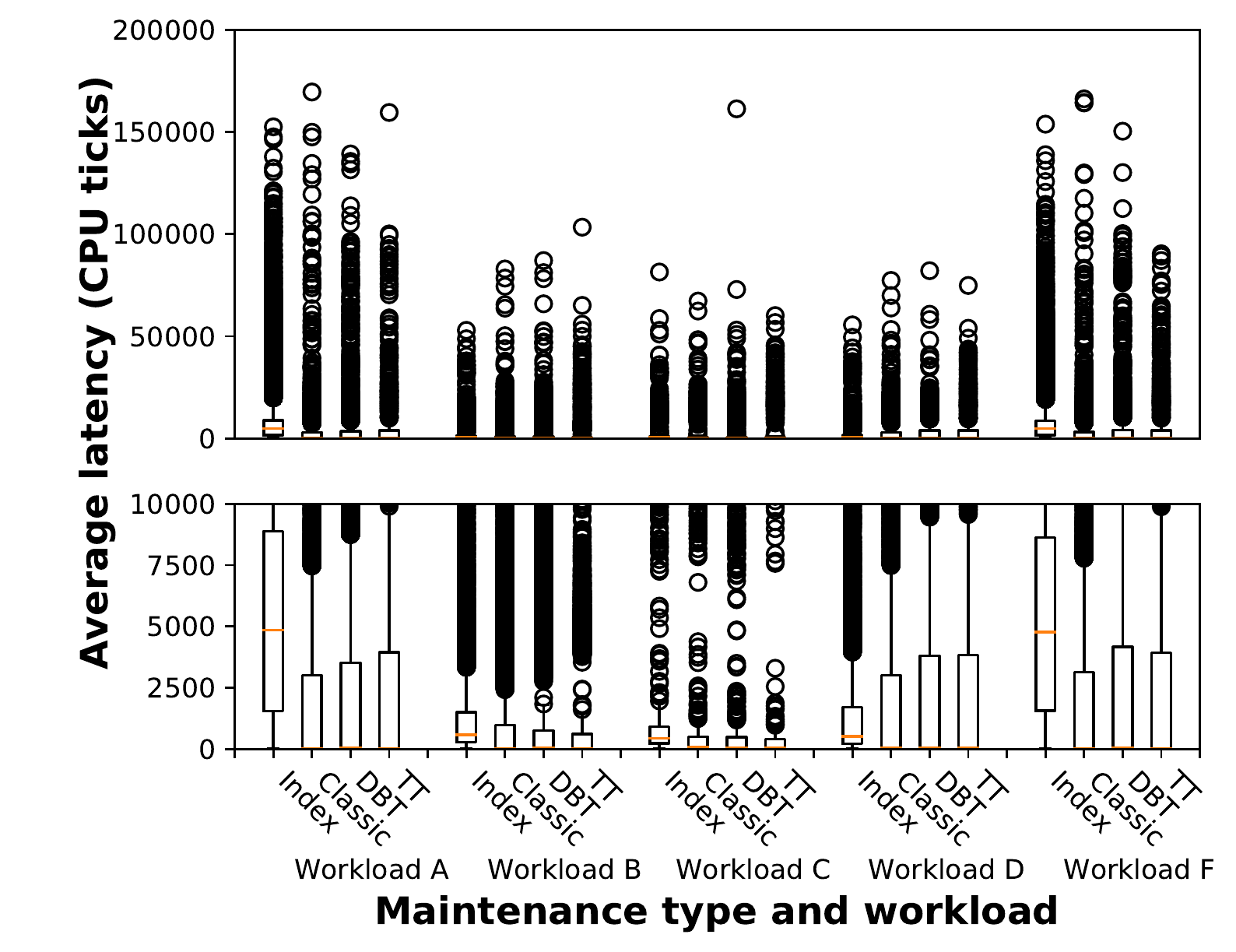}
\caption{Average IVM Operational Latency.  The bottom plot is a zoomed-in view of the top plot.}
\label{fig:stacked_plot}
\trimfigurespacing
\end{figure}

\subsection{Workload}
\label{sssec:jitdrules}

\revised{
To evaluate \systemname, we rely on a benchmark workload created by \jitd~\cite{balakrishnan2019just,kennedy2015just}, an index designed like a just-in-time compiler.
\jitd's underlying data structure is modeled after an AST, allowing a JIT runtime to incrementally and asynchronously rewrite it in the background using pattern-replacement rules~\cite{DBLP:conf/dbpl/BalakrishnanZK19} to support more efficient reads.  
Data is organized through 5 node types that closely mimic the building blocks of typical index structures:
\vspace*{-2mm}
\begin{align*}
&\NodeExplicit{\texttt{Array}}{\texttt{data:Seq[<key:Int,value:Int>]}}{\emptyset}\\[-1mm]
&\NodeExplicit{\texttt{Singleton}}{\texttt{data:<key:Int,value:Int>}}{\emptyset}\\[-1mm]
&\NodeExplicit{\texttt{DeleteSingleton}}{\texttt{key:Int}}{\Node_1}\\[-1mm]
&\NodeExplicit{\texttt{Concat}}{\emptyset}{\Node_1,\Node_2}\\[-1mm]
&\NodeExplicit{\texttt{BinTree}}{\texttt{sep:Int}}{\Node_1,\Node_2}
\vspace*{-7mm}
\end{align*}
}
\revised{
\jitd was configured to use five pattern-replacement rules that mimic Database Cracking~\cite{DBLP:conf/cidr/IdreosKM07} by incrementally building a tree, while pushing updates (\texttt{Singleton} and \texttt{DeleteSingleton} respectively) down into the tree.
}

\noindent \revised{
\textbf{CrackArray}: This rule matches \texttt{Array} nodes and partitions them on a randomly selected pivot $\texttt{sep} \in \texttt{data}$.
\vspace*{-3mm}
\begin{multline*}
\PatternNode{\texttt{Array}}{[\texttt{data}]}{\emptyset}{\True} \rightarrow
\GenerateNode{\texttt{BinTree}}{[\texttt{sep}]}{[\\
  \GenerateNode{\texttt{Array}}{[\comprehension{x}{x.\texttt{key} < \texttt{sep}}]}{[]},\\
  \GenerateNode{\texttt{Array}}{[\comprehension{x}{x.\texttt{key} \geq \texttt{sep}}]}{[]},
]}
\end{multline*}
}
\vspace*{-3mm}

\noindent \revised{
\textbf{PushDownSingletonBtreeLeft/Right}: These rules push \texttt{Singleton} nodes down into \texttt{BinTree} depending on the $\texttt{sep}$arator. 
\vspace*{-3mm}
\begin{multline*}
\PatternNode{\texttt{Concat}}{C}{[
  \PatternNode{\texttt{BinTree}}{B}{\Pattern_1,\Pattern_2}{\emptyset},\\
  \PatternNode{\texttt{Singleton}}{S}{\emptyset}{\emptyset}]}{S.\texttt{key} < \texttt{sep}}
  \rightarrow \\
    \GenerateNode{\texttt{BinTree}}{[\texttt{sep}]}{[
      \GenerateNode{\texttt{Concat}}{[]}{[\\
        \GenerateExisting{\Pattern_1},
        \GenerateExisting{S},
      ]},
      \GenerateExisting{\Pattern_2}, 
    ]}
\end{multline*}
PushDownSingletonBtreeRight is defined analogously.
}

\noindent \revised{
\textbf{PushDownDeleteSingletonBtreeLeft/Right}: These rules push  \texttt{DeleteSingleton} nodes depending on the $\texttt{sep}$arator and are defined analogously to PushDownSingletonBtreeLeft.
}

\revised{
  Although these rewrite rules appear relatively simple, their pattern structures are representative of the vast majority of optimizer in both Apache Spark~\cite{DBLP:conf/sigmod/ArmbrustXLHLBMK15} and Orca~\cite{DBLP:conf/sigmod/SolimanAREGSCGRPWNKB14}.
  A detailed discussion of these rules and how they relate to the example patterns is provided in an accompanying technical report~\cite{balakrishnan2021treetoaster}.
}

\subsection{Data Gathering and Measurement}
\revised{
We instrumented \jitd to collect updates to AST nodes as allocation (\texttt{insert()}) and garbage collection (\texttt{remove()}) operations.
To vary the distribution of optimization opportunities we used the six baseline YCSB~\cite{cooper2010benchmarking} benchmark workloads as input to \jitd.
Each workload exercises a different set of node operations, resulting in ASTs composed of different node structures, patterns, and the applicability of different rewrite rules.
We built a testing module in C++, allowing us to replace \jitd's naive tree traversal with the view maintenance schemes described above for an apples-to-apples comparison.
\Cref{fig:benchmark_system} illustrates the benchmark generation process.
}

Views for \systemname and label indexing were generated by declarative specification as described in \Cref{sec:inlining} and views for \dbt were generated by hand, translating rules to equivalent SQL as described in \Cref{sec:background}.

We instrumented \systemname to collect in-structure information pertaining to view materialization and maintenance:  the time to identify potential \jitd transform operations (rows in the materialized views), and the time to maintain the views in response to updates.
The test harness also records database operation latency and process memory usage, as reported by the Linux \texttt{/proc} interface.
To summarize, we measure performance along three axes: 
(i) Time spent finding a pattern match,
(ii) Time spent maintaining support structures (if any), and 
(iii) Memory allocated.

\subsection{Evaluation}
\jitd is configured to use 5 representative rewrite rules listed above.
Detailed results are grouped by the triggering rule.
Each combination was run 10 times, with the search and operation results aggregated.
Experiments were run on Ubuntu 16.04.06 LTS with 192GB RAM and 24 core Xeon 2.50GHz processors. \rerevised{All the results are obtained from the instrumented \jitd compiler run on YCSB workloads with 300M keys.}

\subsection{Results}
\begin{figure}
\centering
\includegraphics[width=\columnwidth]{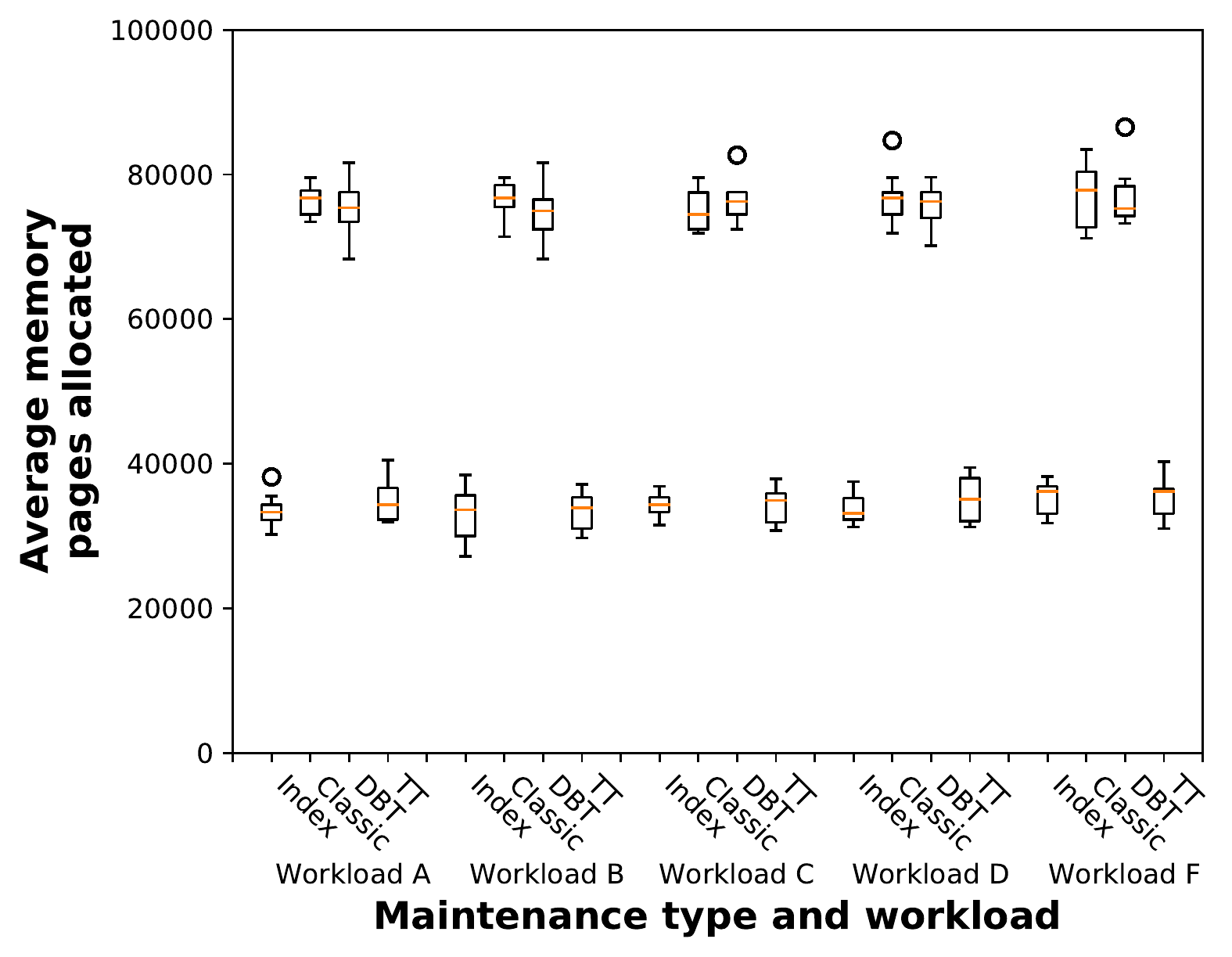}
\caption{Average Process Memory Usage Summary.}
\label{fig:total_memory}
\end{figure}



\rerevised{We first evaluate how IVM performs relative to other methods of IVM in identifying target nodes in tree structure.
We compare the latency in identifying potential nodes (i.e. materializing views) using the 5 IVM methods.}

\rerevised{Figure~\ref{fig:search_performance_all} shows
there were 5 sets of views one per transform (e.g. CrackArray) that were materialized, representing target nodes in the underlying tree structure.}
The 5 boxplot clusters compare the relative average latency of identifying 1 such node, using each of 5 identification methods.
In each case, the naive search approach exhibits the worst performance.
The label index approach also yields worse results than either of the IVM approaches.
For identifying target nodes, we conclude that an IVM approach performs better.

We compare \systemname to IVM alternatives, including
label indexing as well as a classic IVM system and a hash-based IVM implemented using \dbt.
\rerevised{Figure~\ref{fig:crossplot} shows
for each system, the graph shows both memory and overall performance, in terms of the combined access latency and search costs per optimizer iteration.}1
\systemname easily outperforms naive iteration and has an advantage over the label index, with only a slight memory penalty relative to the former.
It slightly outperforms \dbt in general, but the total system memory for \systemname is less than half of \dbt.

Figure~\ref{fig:combination_performance_all} shows the average total latency spent searching for a target node for a \jitd reorganization step, plus all maintenance steps in the reorganization, for each of the 4 IVM target node identification methods.
\rerevised{While the label-index approach performs well on workloads where the structure is allowed to converge (loads B and D) it scales poorly under increased pressure (update heavy loads A and F). Average total time was significantly worse than that of \systemname.
In terms of total cost, \systemname outperforms both classic IVM and \dbt IVM.}


Finally, Figure~\ref{fig:total_memory} shows the average memory use in pages.
For all the methods, the memory footprint was a relatively stable constant within each run, with only small inter-run variance.
Comparing across materialization and maintenance methods, both classic IVM and \dbt exhibited significantly greater memory consumption, an expected result due to its strategy of maintaining large pre-computed tables.
Despite using significantly less memory to optimize performance, \systemname performs as well as if not significantly better than these 2 alternatives.
Figure~\ref{fig:stacked_plot} shows an aggregate summary of all workloads.
Overall, \systemname offers both better memory and latency across all alternatives.




\section{Related Work}
\label{sec:related}

\systemname builds on decades of work in Incremental View Maintenance (IVM) --- See \cite{DBLP:conf/sigmod/ColbyGLMT96} for a survey.
The area has been extensively studied, with techniques developed for incremental maintenance support for of a wide range of data models~\cite{DBLP:conf/sigmod/RossSS96,DBLP:conf/sigmod/BlakeleyLT86,DBLP:conf/icde/ChaudhuriKPS95,DBLP:journals/vldb/YangW03} and query language features~\cite{DBLP:conf/vldb/PalpanasSCP02,DBLP:conf/dbpl/KawaguchiLMR97,DBLP:conf/pods/Koch10,DBLP:conf/icde/LarsonZ07}.

Notable are techniques that improve performance through dynamic programming~\cite{DBLP:conf/sigmod/RossSS96,DBLP:journals/vldb/KochAKNNLS14,DBLP:conf/edbt/YangK17,DBLP:conf/cidr/McSherryMII13}.  
A common approach is materializing intermediate results; For one plan as proposed by Ross et. al.~\cite{DBLP:conf/sigmod/RossSS96}, or all possible plans as proposed by Koch et. al~\cite{DBLP:journals/vldb/KochAKNNLS14}.  
A key feature of both approaches is computing the mininmal update -- or slice -- of the query result, an idea core to systems like Differential Dataflow~\cite{DBLP:conf/cidr/McSherryMII13}.
Both approaches show significant performance gains on general queries.
However, the sources of these gains: selection-pushdown, aggregate-pushdown, and cache locality are less relevant in the context of abstract syntax trees.
Similarly-spirited approaches can be found in other contexts, including graphical inference~\cite{DBLP:conf/edbt/YangK17}, and fixed point computations~\cite{DBLP:conf/cidr/McSherryMII13}.  

Also relevant is the idea of embedding query processing logic into a compiled application~\cite{DBLP:conf/sigmod/MeijerBB06,sqlite,DBLP:conf/sigmod/RaasveldtM19,ahmad2012dbtoaster,DBLP:journals/vldb/KochAKNNLS14,DBLP:conf/sigmod/NikolicD016,berkeleydb,Shaikhha:213124}.
Systems like BerkeleyDB, SQLite, and DuckDB embed full query processing logic, while systems like DBToaster~\cite{ahmad2012dbtoaster,DBLP:journals/vldb/KochAKNNLS14,DBLP:conf/sigmod/NikolicD016} and LinQ~\cite{DBLP:conf/sigmod/MeijerBB06} compile queries along with the application, making it possible to generate native code optimized for the application.
Most notably, this makes it possible to aggressively inline SQL and imperative logic, often avoiding the need for boxing, expensive VM transitions for user-defined functions, and more~\cite{Shaikhha:213124,DBLP:conf/sigmod/MeijerBB06,DBLP:conf/aosd/Wurthinger14}.  
Major database engines have also recently been extended to compile queries to native code~\cite{DBLP:journals/pvldb/Neumann11,DBLP:journals/pvldb/ButtersteinG16,tungsten}, albeit at query compile time.

To our knowledge, IVM over Abstract Syntax Trees has not been studied directly.
\revised{
The Cascades framework~\cite{DBLP:journals/debu/Graefe95a} considers streamlined approaches to scheduling rule application, a strategy that is used by the Orca~\cite{DBLP:conf/sigmod/SolimanAREGSCGRPWNKB14} compiler.
}
IVM approaches also exist for general tree and graph query languages and data models like XPath~\cite{DBLP:conf/er/DimitrovaER03}, Cypher~\cite{DBLP:conf/sigmod/Szarnyas18,DBLP:journals/corr/abs-1806-07344}, and the Nested Relational Calculus~\cite{DBLP:conf/pods/0001LT16}.
These schemes address recursion, with which join widths are no longer bounded; and aggregates without an abelian group representation (e.g., min/max), where deletions are more challenging.

However, two approaches aimed at the object exchange model~\cite{DBLP:conf/vldb/AbiteboulMRVW98,DBLP:conf/icde/ZhugeG98}, are very closely related to our own approach.
One approach proposed by Abiteboul et. al.~\cite{DBLP:conf/vldb/AbiteboulMRVW98} first determines the potential positions at which an update could affect a view and then uses the update to recompute the remaining query fragments.
However, its more expressive query language limits optimization opportunities, creating situations where it may be faster to simply recompute the view query from scratch.
The other approach proposed by Zhuge and Molina~\cite{DBLP:conf/icde/ZhugeG98} follows a similar model to our immutable IVM scheme, enforcing pattern matches on ancestors by recursively triggering shadow updates to all ancestors.

\section{Conclusion and Future Work} 
In this paper we introduce a formalized mechanism for
pattern-matching queries over ASTs and IVM over such queries. 
Our realization of the theory in the just-in-time data-structure compiler shows that the \systemname approach works.

\rerevised{
Many compilers choose to specify rewrites over the AST as pattern match rules, whether declaratively through language features (e.g., Spark), or a customized implementation of the same (e.g., Orca).
We would like to extend our work, integrating it with language-based pattern matching to automate the matching process.
For example, a DSL implemented with Scala macros could extract declarative pattern matches, plug them into our grammar, and provide a nearly drop-in replacement for its existing optimizer infrastructure.
}
\rerevised{We also plan on extending our approach to work with graphs. This will allow for recursive pattern matches and efficient IVM over graph structures. This is particularly interesting for traditional compilers as there exist many optimizations that rely on fixed-points while traversing control-flow graphs. }

\begin{acks}
This work is supported by NSF grants: SHF-1749539, IIS-1617586, and IIS-1750460.  All opinions presented are those of the authors.  The authors wish to thank the reviewers and shepherd for their substantial contributions.
\end{acks}


\bibliographystyle{ACM-Reference-Format}
\bibliography{main}
\balance

\pagebreak
\IfTechReport{\appendix


\begin{figure}
  \centering
  \begin{subfigure}{0.45\textwidth}
    \centering
    \includegraphics[width=0.8\textwidth]{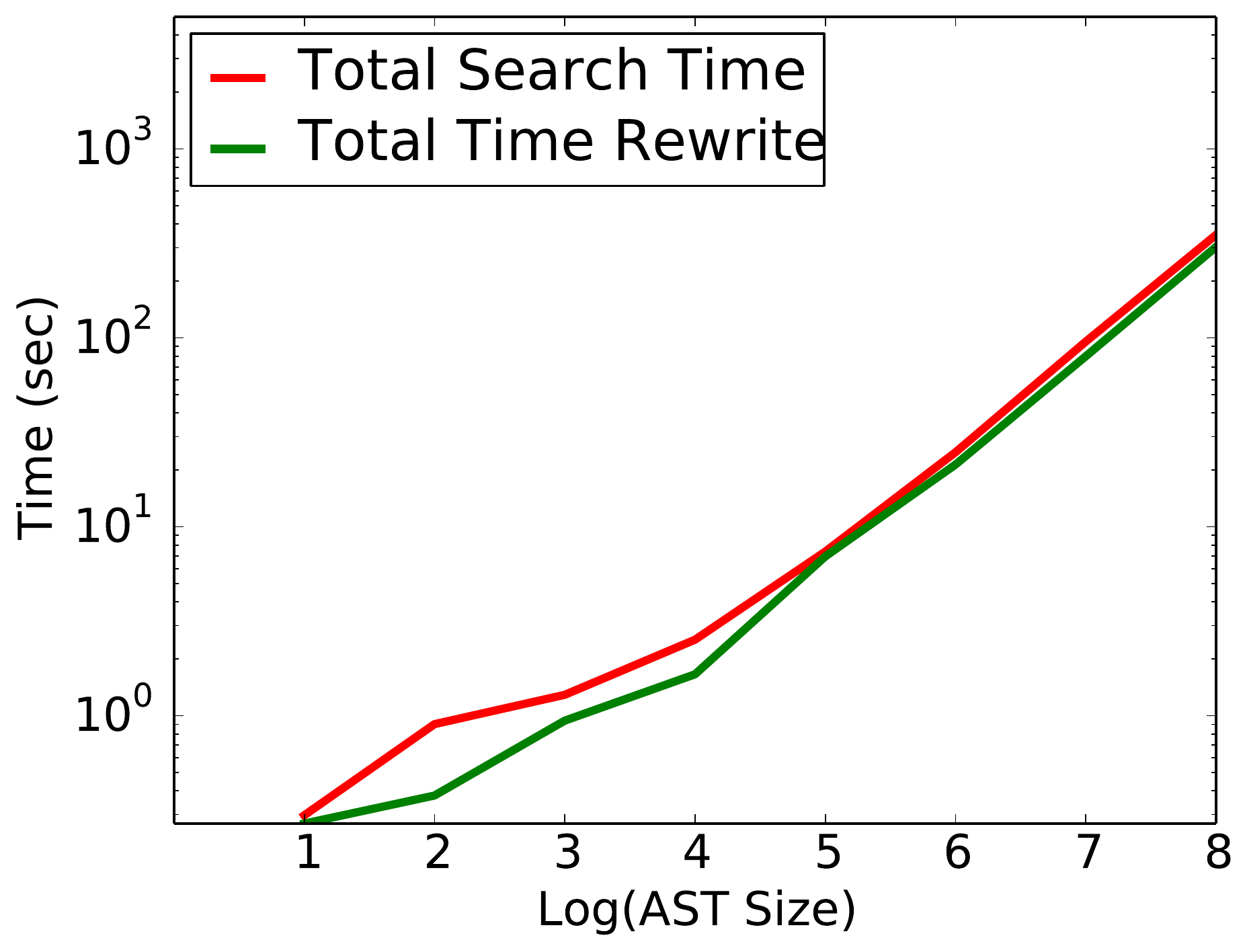}
    \caption{Spark: Total time spent in rewrite}
    \label{fig:SparkTotal}
  \end{subfigure}
  \begin{subfigure}{0.45\textwidth}
    \centering
    \includegraphics[width=0.8\textwidth]{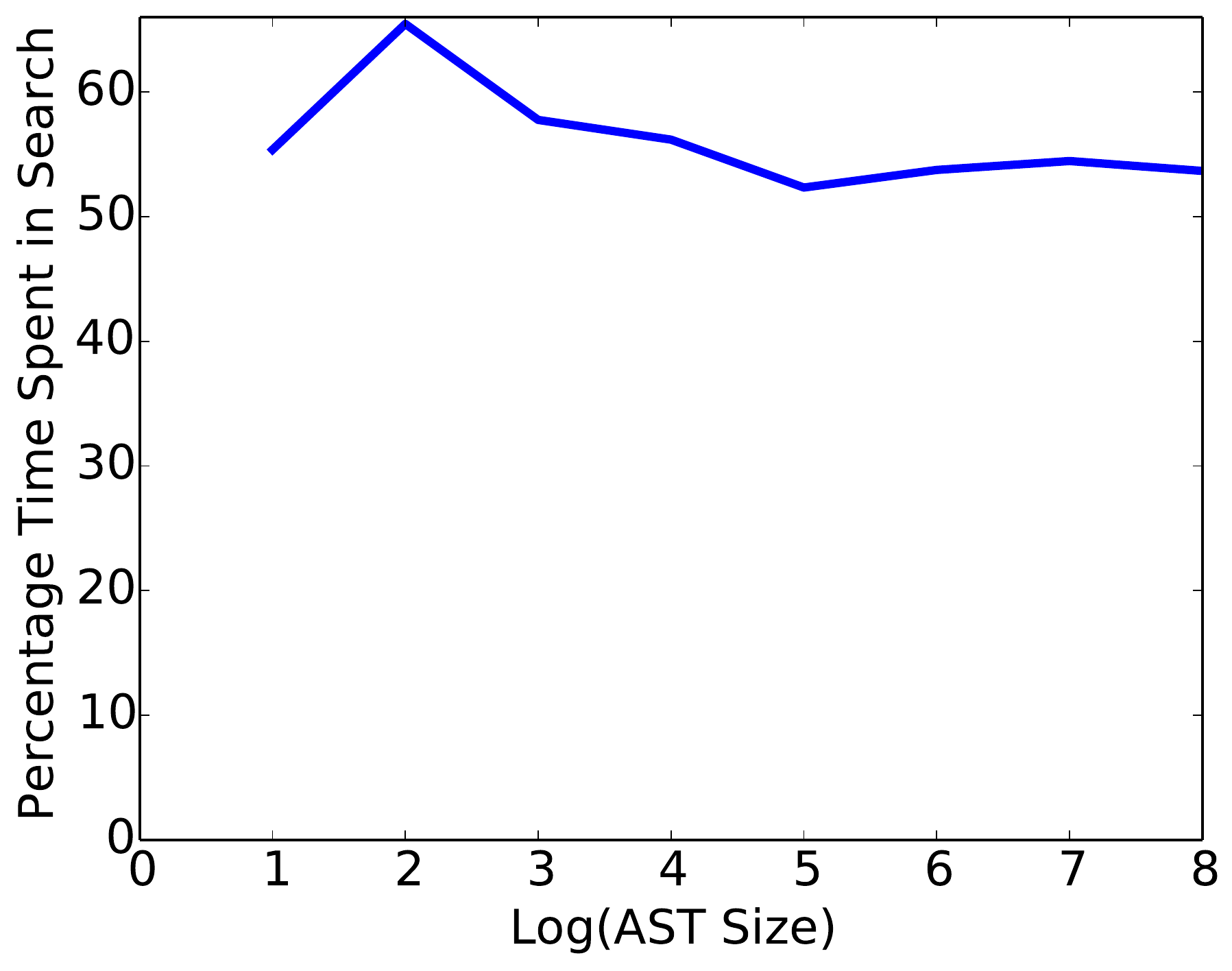}
    \caption{Spark: Percentage time in AST search.}
    \label{fig:SparkFraction}
  \end{subfigure}
  \caption{Rewrite and Search Times for Spark's Optimizer}
\end{figure}

\section{Exploration of Spark and ORCA}
In this section we present a thorough exploration of Spark's Catalyst optimizer and Greenplum's Orca. We have studied two open source SQL optimizers with an eye towards understanding how much of a time sink AST searches are.  To accomplish this we used a simple, easily scalable query pattern:
\begin{lstlisting}[style=psqlcolor]
CREATE VIEW TABLE_[N] AS
  SELECT * FROM (    SELECT * FROM TABLE_[N-1] 
           UNION ALL SELECT * FROM TABLE_[N-1]
  ) a, 
  SELECT * FROM (    SELECT * FROM TABLE_[N-1] 
           UNION ALL SELECT * FROM TABLE_[N-1]
  ) b,
  WHERE a.attr = b.attr
\end{lstlisting}
This query structure is representative of an antipattern that arises naturally in query rewriting for provenance-tracking (e.g., \cite{vizier,gprom}), and that such compilers must explicitly guard against. 
Results appear below in \Cref{fig:OrcaTotal,fig:OrcaFraction,fig:SparkTotal,fig:SparkFraction}.
Spark results shown are the average of 5 runs.  
Orca timings were noisier, so we take the average of 10 runs.
Spark's optimizer works through Scala's native pattern matching syntax (a recursive \texttt{match \{ case \}} blocks, or more precisely calls to Spark's \texttt{transform \{ case \}} utility function).  
We obtained these timings by logging the time spent in each state i.e, serach for a pattern and apply a pattern.
Orca's compiler has a more intricate rule scheduling mechanism, but also works by recursive tree traversal during which a pairwise recursive traversal of the pattern AST and AST subtrees is used to check for matches.  We measure the time taken in this match and contrast.

\textbf{Take-away:} Both Catalyst and Orca as seen in \Cref{fig:SparkFraction,fig:OrcaFraction} spend a non-negligible fraction of their optimization time searching for candidate AST nodes to rewrite (50-60\% for Spark, 5-20\% for Orca).  In both cases, the relative fraction of time spent searching drops asymptotically (to about 50\%, 5\% respectively) as the AST size grows, but continues \textbf{scaling linearly with the AST size} \Cref{fig:SparkTotal,fig:OrcaTotal}.
These results suggest that (i) At small scales, pattern-matching is a dominant cost for query optimization, and (ii) Any comprehensive strategy that will allow optimizers to scale to enormous ASTs will need to include a technique analogous to \systemname.

\section{JITD Compiler, Spark and ORCA}
We did a thorough assessment of optimizer rules in Catalyst and ORCA, and found that most of the rules were comparable to, if not strict subsets of the rules used in the \jitd compiler we were evaluating --- and all of these rules are local patterns. We first represent Spark and ORCA AST Nodes with our grammar in \Cref{sec:nodes} and present a detailed overview of our assessment results in \Cref{sec:spark transforms,sec: orca transforms}.

\begin{figure}
  \centering
  \begin{subfigure}{0.45\textwidth}
    \centering
    \includegraphics[width=0.8\textwidth]{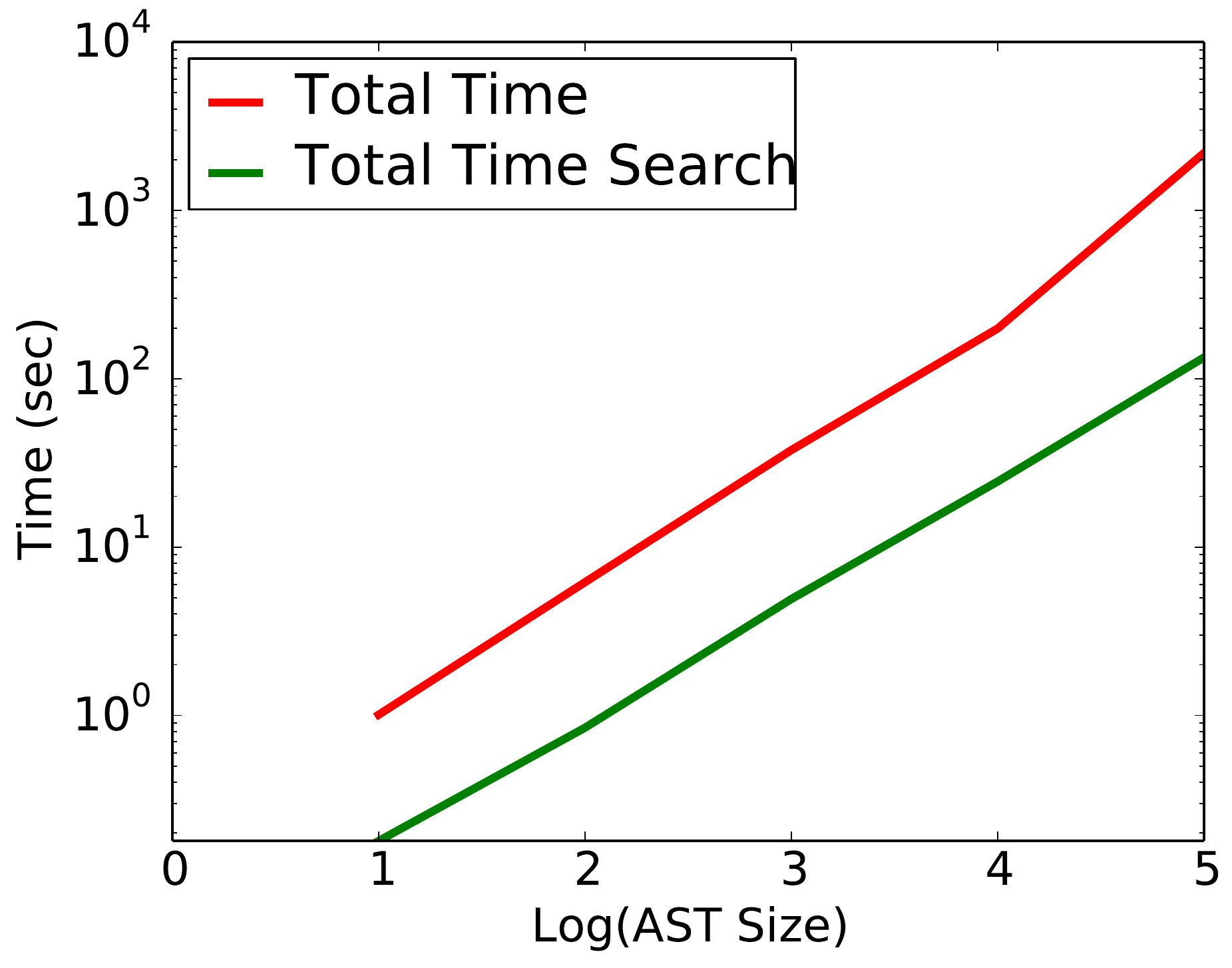}
    \caption{Orca: Total time spent in rewrite}
    \label{fig:OrcaTotal}
  \end{subfigure}
  \begin{subfigure}{0.45\textwidth}
    \centering
    \includegraphics[width=0.8\textwidth]{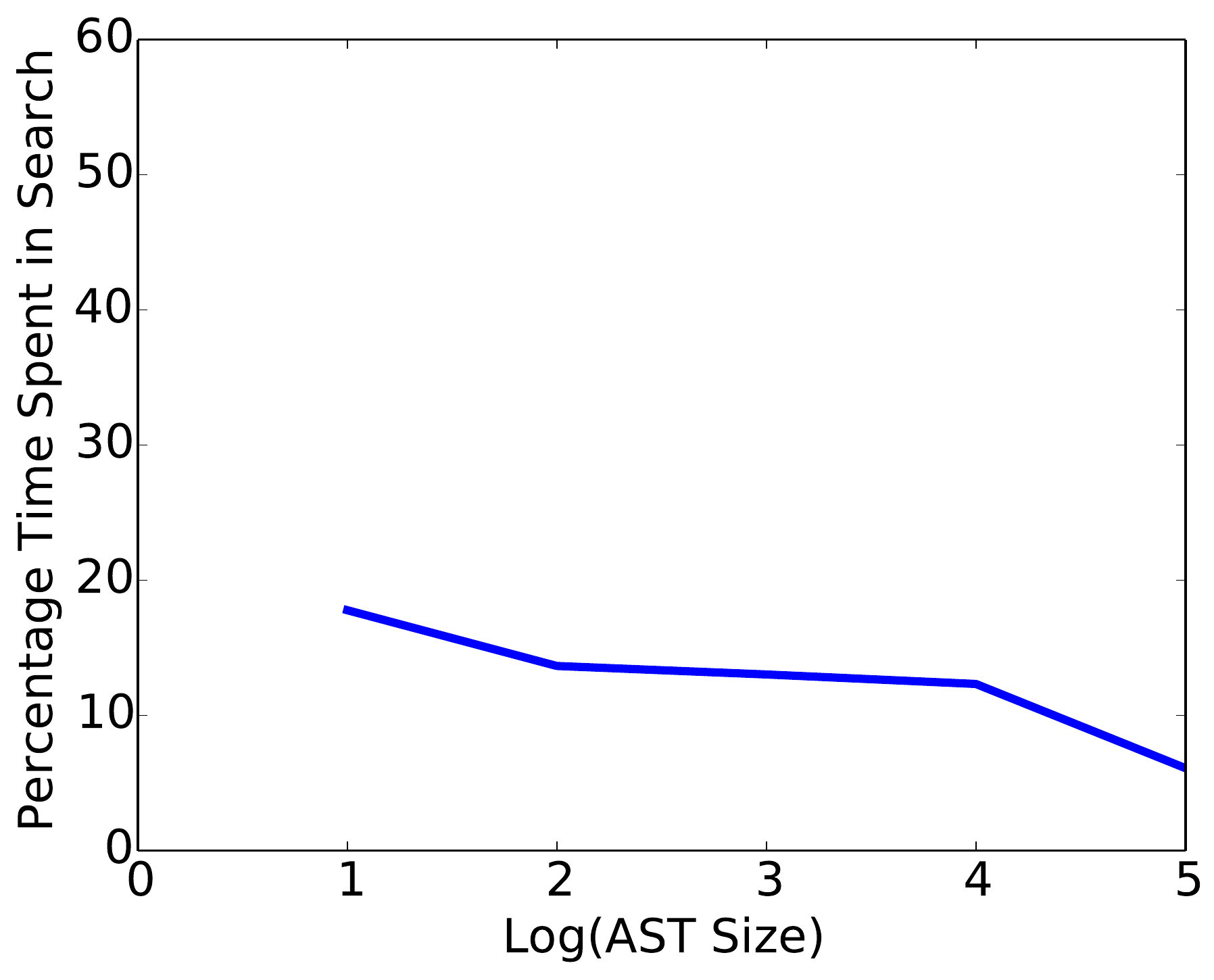}
    \caption{Orca: Percentage time in AST search.}
    \label{fig:OrcaFraction}
  \end{subfigure}
  \caption{Rewrite and Search Times for Spark's Optimizer}
\end{figure}

\section{Spark and ORCA AST Schemas}
\label{sec:nodes}
This subsection defines a simplified schema (in terms of the node's attributes and children) for nodes in Spark's \texttt{LogicalPlan} AST and ORCA's \texttt{CExpression} AST.  Because all nodes in Spark share common attributes, we also allow pattern matching on a node with a variable label $\ell$:
$$\NodeExplicit{\Label}{\texttt{[o:Seq[Attribute],r:AttributeSet,\ldots]}}{\NodeChildren}$$

\begin{figure*}
\begin{lstlisting}
SubqueryExpr{[o:Seq[Attribute],r:AttributeSet,ce:Seq[Expression],e$_{id}$:ExprId]}{$\Node_1$}
UnaryNode{[o:Seq[Attribute],r:AttributeSet,e: Seq[Expression,..]}{$\Node_1$}
Project{[o:Seq[Attribute],r:AttributeSet,p:Seq[NamedExpr]]}{$\Node_1$}
Window{[o:Seq[Attribute],r:AttributeSet,w:Seq[NamedExpr],ps:Seq[Expr],...]}{$\Node_1$}
Filter{[o:Seq[Attribute],r:AttributeSet,c:Expression]}{$\Node_1$}
Aggregate{[o:Seq[Attribute],r:AttributeSet,g:Seq[Expr],a:Seq[NamedExpr]]}{$\Node_1$}
Union{[o:Seq[Attribute],r:AttributeSet,p:Bool,a:Bool]}{$\NodeList$}
WaterMark{[o:Seq[Attribute],r:AttributeSet,e:EventTime,d:Delay]]}{$\Node_1$}
Join{[o:Seq[Attribute],r:AttributeSet,j:JoinType,c:Expression,h:JoinHint]}{$\Node_1$,$\Node_2$}
Expand{[o:Seq[Attribute],r:AttributeSet,p:Seq[Seq[Expr]],op:Seq[Attribute]]}{$\Node_1$}
Deserialize{[o:Seq[Attribute],r:AttributeSet,e:Expr,a:Attr]}{$\Node_1$}
FlatMap{[o:Seq[Attribute],r:AttributeSet,e:Expr,a:Attr]}{$\Node_1$}
ScriptTransf{[o:Seq[Attribute],r:AttributeSet,i:Seq[Expr],s:String,...]}{$\Node_1$}
Distinct{[o:Seq[Attribute],r:AttributeSet]}{$\Node_1$}
Generate{[o:Seq[Attribute],r:AttributeSet,g: Generator, u: Seq[Int], ...]}{$\Node_1$}
SetOp{[o:Seq[Attribute],r:AttributeSet]}{$\Node_1$,$\Node_2$}
Subquery{[o:Seq[Attribute],r:AttributeSet,c:Bool]}{$\Node_1$}
Limit{[o:Seq[Attribute],r:AttributeSet,c:Expression]}{$\Node_1$}
LocalRel{[o:Seq[Attribute],r:AttributeSet,op:Seq[Attribute],...]}{}
Repartion{[o:Seq[Attribute],r:AttributeSet,n:Int,s:Bool]}{$\Node_1$}
GlobalLimit{[o:Seq[Attribute],r:AttributeSet,g:Expr]]}{$\Node_1$}
LocalLimit{[o:Seq[Attribute],r:AttributeSet,l:Expr]]}{$\Node_1$}
Sample{[o:Seq[Attribute],r:AttributeSet,l: Double, u: Double, w: Bool, ...]}{$\Node_1$}
\end{lstlisting}
\label{fig:sparknodes}
\caption{A selection of Spark's \texttt{LogicalPlan} node types expressed as schemas for $\mathcal G$}
\end{figure*}

\bigskip


\begin{figure*}
\begin{lstlisting}
CLogicalNAryJoin{[exprhdl:CExpressionHandle,p:childPredicateList]}{$\NodeChildren$}
CLogicalGet{[exprhdl:CExpressionHandle,t:CName,pt:CTableDescriptor,...]}{$\emptyset$}
CLogicalSelect{[exprhdl:CExpressionHandle,m:CHashMapExpr2Expr,
        pt:CTableDescriptor,predicateExpr:CExpression]}{$\Node_1$}
CLogicalInnerJoin
        {[exprhdl:CExpressionHandle,predicateExpr:CExpression]}{$\Node_1$,$\Node_2$}
CLogicalUnionAll 
        {[exprhdl:CExpressionHandle,i:Int,o:CColRefArray,k:CColRef2dArray]}{$\NodeChildren$}
\end{lstlisting}
\label{fig:orcanodes}
\caption{A selection of Orca's AST node types expressed as schemas for $\mathcal G$}
\end{figure*}


\section{Spark Transforms}
Scala's optimizer makes extensive use of \texttt{LogicalPlan}'s \texttt{transform} method (among several variants), which does a pattern-matching search and replaces AST nodes based on a Scala pattern match (a \texttt{case} clause).  In the following, we provide (i) the \texttt{case} clauses, (ii) The corresponding pattern matching expression in our grammar, and (iii) The most similar transform.

\newcommand{\sparkscala}{\noindent \textbf{Scala:}}
\newcommand{\orcacpp}{\noindent \textbf{Cpp:}}
\newcommand{\grammarpattern}{\noindent \textbf{Patterns:}\\}
\newcommand{\similartojitd}[1]{\noindent \textbf{Most Similar JITD Pattern:} #1}

\newcommand{\formatpatternmatch}[1]{\noindent $#1$ \\}

\label{sec:spark transforms}
\subsection{\textbf{Transform:} RemoveNoopOperators}
\sparkscala
\begin{lstlisting}[style=scala]
case p : Project(_, child) if child.sameOutput(p)
case w: Window if w.windowExpressions.isEmpty
\end{lstlisting}

\grammarpattern
\formatpatternmatch{\PatternNode{\texttt{Project}}{\texttt{[o$_1$,r,p]}}{\PatternNode{\Label}{\texttt{[o$_2$,\ldots]}}{\PatternChildren}{\True}}{\\\hspace*{20mm}\texttt{o$_1$}=\texttt{o$_2$}}}
\formatpatternmatch{\PatternNode{\texttt{Window}}{\texttt{[o,r,w,ps,so]}}{\Pattern_1}{\texttt{w.empty}}}

\similartojitd{DeleteSingletonFromArray}


\subsection{\textbf{Transform:} CombineFilters}
\sparkscala
\begin{lstlisting}[style=scala]
case Filter(fc, nf : Filter(nc, grandChild)) 
         if fc.deterministic && nc.deterministic
\end{lstlisting}

\grammarpattern
\formatpatternmatch{
  \PatternNode{\texttt{Filter}}{\texttt{[o$_1$,r$_1$,c$_1$]}}{\\\hspace*{20mm}\PatternNode{\texttt{Filter}}{\texttt{[o$_2$,r$_2$,c$_2$]}}{\Pattern_1}{\texttt{c$_2$.deterministic}}}{\\\hspace*{10mm}\texttt{c$_1$.deterministic}}
}

\similartojitd{MergeSortedConcat,PushDownDowntDeleteSingletonLeft/Right}

\subsection{\textbf{Transform:} PushPredicateThroughNonJoin}
\sparkscala
\begin{lstlisting}[style=scala]
case Filter(condition, project : 
        Project(fields,grandChild)) 
            if fields.forall(_.deterministic) && 
                canPushThroughCondition(grandChild,
                condition)
\end{lstlisting}
\begin{lstlisting}[style=scala]
case filter : Filter(condition,
        aggregate: Aggregate)
        if aggregate.aggregateExpressions.forall(
            _.deterministic) &&
            aggregate.groupingExpressions.nonEmpty 
\end{lstlisting}
\begin{lstlisting}[style=scala]
case filter : Filter(condition, w: Window)
        if w.partitionSpec.forall(
            _.isInstanceOf[AttributeReference])
\end{lstlisting}
\begin{lstlisting}[style=scala]
case filter : Filter(condition, union: Union)
\end{lstlisting}
\begin{lstlisting}[style=scala]
case filter : Filter(condition, watermark: 
        EventTimeWatermark)
\end{lstlisting}
\begin{lstlisting}[style=scala]
case filter : Filter(_, u: UnaryNode)
    if canPushThrough(u) && u.expressions.forall(
        _.deterministic)
\end{lstlisting}
\begin{lstlisting}[style=scala]
def canPushThrough(p: UnaryNode): Boolean =
p match {
    case _: AppendColumns => true
    case _: Distinct => true
    case _: Generate => true
    case _: Pivot => true
    case _: RepartitionByExpression => true
    case _: Repartition => true
    case _: ScriptTransformation => true
    case _: Sort => true
    case _: BatchEvalPython => true
    case _: ArrowEvalPython => true
    case _: Expand => true
    case _ => false
  }
\end{lstlisting}

\grammarpattern
\formatpatternmatch{
\PatternNode{\texttt{Filter}}{\texttt{[o$_1$,r$_1$,c]}}{
  \PatternNode{\texttt{Project}}{\texttt{[o$_2$,r$_2$,p]}}{
    \\\hspace*{20mm}\PatternNode{\Label}{\texttt{[o$_3$,r$_3$,\ldots]}}{\PatternChildren}{\True}}{\\\hspace*{20mm}\texttt{p.deterministic}}}
    {\\\hspace*{20mm}\texttt{canPushThroughCondition(c},\\\hspace*{20mm}\PatternNode{\Label}{\texttt{[o$_3$,r$_3$,\ldots]}}{\PatternChildren}{\True})}
}
\formatpatternmatch{\PatternNode{\texttt{Filter}}{\texttt{[o$_1$,r$_1$,c]}}{\\\hspace*{20mm}\PatternNode{\texttt{Aggregate}}{\texttt{[o$_2$,r$_2$,g,a]}}{\Pattern_1}{\True}}{\\\hspace*{20mm}\texttt{[a.deterministic,g.empty]}}}
\formatpatternmatch{
\PatternNode{\texttt{Filter}}{\texttt{[o$_1$,r$_1$,c]}}{\\\hspace*{20mm}\PatternNode{\texttt{Window}}{\texttt{[o$_2$,r$_2$,w,ps,so]}}{\Pattern_1}{\\\hspace*{10mm}\texttt{ps.isInstanceOf[AttributeReference]}}}{\True}
}
\formatpatternmatch{
\PatternNode{\texttt{Filter}}{\texttt{[o$_1$,r$_1$,c]}}{\\\hspace*{20mm}\PatternNode{\texttt{Union}}{\texttt{[o$_2$,r$_2$,p,a]}}{\\\hspace*{20mm}\PatternList}{\True}}{\True}
}
\formatpatternmatch{
\PatternNode{\texttt{Filter}}{\texttt{[o$_1$,r$_1$,c]}}{\\\hspace*{20mm}\PatternNode{\texttt{WaterMark}}{\texttt{[o$_2$,r$_2$,e,d]}}{\\\hspace*{20mm}\Pattern_1}{\True}}{\True}
}
\formatpatternmatch{
\PatternNode{\texttt{Filter}}{\texttt{[o$_1$,r$_1$,c]}}{\\\hspace*{20mm}\PatternNode{\texttt{UnaryNode}}{\texttt{[o$_2$,r$_2$,e,\ldots]}}{\Pattern_1}{\\\hspace*{10mm}\texttt{[e.deterministic,canPushThrough(u)]}}}{\True}
}

\similartojitd{MergeUnSortedConcatArray, PushDownDontDeleteSingletonLeft/Right}

\subsection{\textbf{Transform:} PushPredicateThroughJoin}
\sparkscala
\begin{lstlisting}[style=scala]
case f : Filter(filterCondition, 
        Join(left, right, joinType, 
        joinCondition, hint))
            if canPushThrough(joinType)
\end{lstlisting}
\begin{lstlisting}[style=scala]
case j : Join(left, right, joinType, 
        joinCondition, hint) 
            if canPushThrough(joinType)
\end{lstlisting}
\begin{lstlisting}[style=scala]
private def canPushThrough(joinType: JoinType):
Boolean = joinType match {
    case _: InnerLike | LeftSemi | 
            RightOuter | LeftOuter | LeftAnti |
            ExistenceJoin(_) => true
    case _ => false
  }
\end{lstlisting}

\grammarpattern
\formatpatternmatch{
\PatternNode{\texttt{Filter}}{\texttt{[o$_1$,r$_1$,c$_1$]}}{\\\hspace*{20mm}\PatternNode{\texttt{Join}}{\texttt{[o$_2$,r$_2$,j,c$_2$,h]}}{\Pattern_1,\Pattern_2}{\\\hspace*{20mm}\texttt{canPushThrough(j)}}}{\True}
}
\formatpatternmatch{
\PatternNode{\texttt{Join}}{\texttt{[o,r,j,c,h]}}{\Pattern_1,\Pattern_2}{\texttt{canPushThrough(j)}}
}
\similartojitd{
PushDownAndCrack, MergeSortedBTrees}

\subsection{\textbf{Transform:} ColumnPruning}
\sparkscala
\begin{lstlisting}[style=scala]
case p : Project(_, p2: Project) 
        if !p2.outputSet.subsetOf(p.references) 
case p : Project(_, a: Aggregate) 
        if !a.outputSet.subsetOf(p.references)
case a : Project(_, e : Expand(_, _, grandChild)) 
        if !e.outputSet.subsetOf(a.references)
case d : DeserializeToObject(_, _, child) 
        if !child.outputSet.subsetOf(d.references)
case a : Aggregate(_, _, child)
        if !child.outputSet.subsetOf(a.references)
case f : FlatMapGroupsInPandas(_, _, _, child) 
        if !child.outputSet.subsetOf(f.references)
case e : Expand(_, _, child) 
        if !child.outputSet.subsetOf(e.references)
case s : ScriptTransformation(_, _, _, child, _) 
        if !child.outputSet.subsetOf(s.references)
case p : Project(_, g: Generate) 
        if p.references != g.outputSet
case j : Join(_, right, LeftExistence(_), _, _)
case p : Project(_, _: SetOperation)
case p : Project(_, _: Distinct)
case p : Project(_, u: Union)
case p : Project(_, w: Window) 
        if
        !w.windowOutputSet.subsetOf(p.references)
case p : Project(_, _: LeafNode)
case p : Project(_, child) 
        if !child.isInstanceOf[Project]
case GeneratorNestedColumnAliasing(p)
case NestedColumnAliasing(p)
\end{lstlisting}
\grammarpattern
\formatpatternmatch{
\PatternNode{\texttt{Project}}{\texttt{[o$_1$,r$_1$,p$_1$]}}{\\\hspace*{20mm}\PatternNode{\texttt{Project}}{\texttt{[o$_2$,r$_2$,p$_2$]}}{\Pattern_1}{\True}}{\texttt{o$_2$} \subseteq \texttt{r$_1$}}
}
\formatpatternmatch{
\PatternNode{\texttt{Project}}{\texttt{[o$_1$,r$_1$,p$_1$]}}{\\\hspace*{20mm}\PatternNode{\texttt{Aggregate}}{\texttt{[o$_2$,r$_2$,g,a]}}{\Pattern_1}{\True}}{\\\hspace*{20mm}\texttt{o$_2$} \subseteq \texttt{r$_1$}}
}
\formatpatternmatch{
\PatternNode{\texttt{Project}}{\texttt{[o$_1$,r$_1$,p$_1$]}}{\\\hspace*{20mm}\PatternNode{\texttt{Expand}}{\texttt{[o$_2$,r$_2$,p$_2$,op]}}{\Pattern_1}{\True}}{\\\hspace*{20mm}\texttt{o$_2$} \subseteq \texttt{r$_1$}}
}
\formatpatternmatch{
\PatternNode{\texttt{Deserialize}}{\texttt{[o$_1$,r$_1$,e,a]}}{\\\hspace*{20mm}\PatternNode{\Label}{\texttt{[o$_2$,r$_2$,\ldots]}}{\PatternChildren}{\True}}{\texttt{o$_2$} \subseteq \texttt{r$_1$}}
}
\formatpatternmatch{
\PatternNode{\texttt{Aggregate}}{\texttt{[o$_1$,r$_1$,g,a]}}{\\\hspace*{20mm}\PatternNode{\Label}{\texttt{[o$_2$,r$_2$,\ldots]}}{\PatternChildren}{\True}}{\texttt{o$_2$} \subseteq \texttt{r$_1$}}
}
\formatpatternmatch{
\PatternNode{\texttt{FlatMap}}{\texttt{[o$_1$,r$_1$,e,a]}}{\\\hspace*{20mm}\PatternNode{\Label}{\texttt{[o$_2$,r$_2$,\ldots]}}{\PatternChildren}{\True}}{\texttt{o$_2$} \subseteq \texttt{r$_1$}}
}
\formatpatternmatch{
\PatternNode{\texttt{Expand}}{\texttt{[o$_1$,r$_1$,p,o]}}{\\\hspace*{20mm}\PatternNode{\Label}{\texttt{[o$_2$,r$_2$,\ldots]}}{\PatternChildren}{\True}}{\texttt{o$_2$} \subseteq \texttt{r$_1$}}
}
\formatpatternmatch{
\PatternNode{\texttt{ScriptTransf}}{\texttt{[o$_1$,r$_1$,i,s,op,io]}}{\\\hspace*{20mm}\PatternNode{\Label}{\texttt{[o$_2$,r$_2$,\ldots]}}{\PatternChildren}{\True}}{\texttt{o$_2$} \subseteq \texttt{r$_1$}}
}
\formatpatternmatch{
\PatternNode{\texttt{Project}}{\texttt{[o$_1$,r$_2$,p]}}{\\\hspace*{20mm}\PatternNode{\texttt{Generate}}{\texttt{[o$_2$,r$_2$,g,u,ob,q,go]}}{\\\hspace*{20mm}\Pattern_1}{\True}}{\texttt{r$_1$} \neq \texttt{o$_2$}}
}
\formatpatternmatch{
\PatternNode{\texttt{Join}}{\texttt{[o,r,j,c,h:leftExistence]}}{\Pattern_1,\Pattern_2}{\True}
}
\formatpatternmatch{
\PatternNode{\texttt{Project}}{\texttt{[o$_1$,r$_1$,p]}}{\\\hspace*{20mm}\PatternNode{\texttt{SetOp}}{\texttt{[o$_2$,r$_2$]}}{\Pattern_1,\Pattern_2}{\True}}{\True}
}
\formatpatternmatch{
\PatternNode{\texttt{Project}}{\texttt{[o$_1$,r$_1$,p]}}{\\\hspace*{20mm}\PatternNode{\texttt{Distinct}}{\texttt{[o$_2$,r$_1$]}}{\Pattern_1}{\True}}{\True}
}
\formatpatternmatch{
\PatternNode{\texttt{Project}}{\texttt{[o$_1$,r$_1$,p$_1$]}}{\\\hspace*{20mm}\PatternNode{\texttt{Union}}{\texttt{[o$_2$,r$_2$,p$_2$,a]}}{\\\hspace*{20mm}\PatternList}{\True}}{\True}
}
\formatpatternmatch{
\PatternNode{\texttt{Project}}{\texttt{[o$_1$,r$_1$,p$_1$]}}{\\\hspace*{20mm}\PatternNode{\texttt{Window}}{\texttt{[o$_2$,r$_2$,w,p$_2$]}}{\Pattern_1}{\True}}{\\\hspace*{20mm}\texttt{o$_2$} \subseteq \texttt{r$_2$}}
}
\formatpatternmatch{
\PatternNode{\texttt{Project}}{\texttt{[o$_1$,r$_1$,p$_1$]}}{\\\hspace*{20mm}\PatternNode{\Label}{\texttt{[o$_2$,r$_2$,\ldots]}}{\emptyset}{\True}}{\True}
}
\formatpatternmatch{
\PatternNode{\texttt{Project}}{\texttt{[o$_1$,r$_1$,p$_1$]}}{\\\hspace*{20mm}\PatternNode{\texttt{Project}}{\texttt{[o$_2$,r$_2$,p$_2$]}}{\Pattern_1}{\True}}{\True}
}
\formatpatternmatch{
\PatternNode{\texttt{Project}}{\texttt{[o$_1$,r$_1$,p$_1$]}}{\Node_1}{\True}
}
\formatpatternmatch{
\PatternNode{\texttt{Generate}}{\texttt{[o$_2$,r$_2$,g,u,ob,q,go]}}{\Pattern_1}{\True}
}
\similartojitd{MergeConcatNodes}


\subsection{\textbf{Transform:} RewritePredicateSubquery}
\sparkscala
\begin{lstlisting}[style=scala]
case Filter(condition, child)
\end{lstlisting}
\grammarpattern
\formatpatternmatch{
\PatternNode{\texttt{Filter}}{\texttt{[o,r,c]}}{\Pattern_1}{\True}
}
\similartojitd{DeleteSingletonFromArray}

\subsection{\textbf{Transform:} RemoveRedundantAliases}
\sparkscala
\begin{lstlisting}[style=scala]
case Subquery(child, correlated)
\end{lstlisting}
\begin{lstlisting}[style=scala]
case Join(left, right, joinType, condition, hint)
\end{lstlisting}

\grammarpattern
\formatpatternmatch{
\PatternNode{\texttt{Subquery}}{\texttt{[o,r,c]}}{\Pattern_1}{\True}
}
\formatpatternmatch{
\PatternNode{\texttt{Join}}{\texttt{[o,r,j,c,h]}}{\Pattern_1,\Pattern_2}{\True}
}
\similartojitd{CrackArray,DeleteSingletonFromArray}

\subsection{\textbf{Transform:} InferFiltersFromConstraints}
\sparkscala
\begin{lstlisting}[style=scala]
case filter : Filter(condition, child)
\end{lstlisting}
\begin{lstlisting}[style=scala]
case join : Join(left, right,
                joinType, conditionOpt, _)
\end{lstlisting}
\grammarpattern
\formatpatternmatch{
\PatternNode{\texttt{Filter}}{\texttt{[o,r,c]}}{\Pattern_1}{\True}
}
\formatpatternmatch{
\PatternNode{\texttt{Join}}{\texttt{[o,r,j,c,h]}}{\Pattern_1,\Pattern_2}{\True}
}

\similartojitd{CrackArray,DeleteSingletonFromArray}

\subsection{\textbf{Transform:} ConvertToLocalrelation}
\sparkscala
\begin{lstlisting}[style=scala]
case Project(projectList, 
        LocalRelation(output, data, isStreaming))
        if !projectList.exists(hasUnevaluableExpr)
\end{lstlisting}
\begin{lstlisting}[style=scala]
case Limit(IntegerLiteral(limit), 
        LocalRelation(output, data, isStreaming))
\end{lstlisting}
\begin{lstlisting}[style=scala]
case Filter(condition, 
        LocalRelation(output, data, isStreaming))
            if !hasUnevaluableExpr(condition)
\end{lstlisting}
\begin{lstlisting}[style=scala]
private def hasUnevaluableExpr(expr: Expression): 
Boolean = {
    expr.find(e => e.isInstanceOf[Unevaluable] &&
    !e.isInstanceOf[AttributeReference]).isDefined
 }
\end{lstlisting}
\grammarpattern 
\formatpatternmatch{
\PatternNode{\texttt{Project}}{\texttt{[o$_1$,r$_2$,p]}}{\\\hspace*{20mm}\PatternNode{\texttt{LocalRelation}}{\texttt{[o$_2$,r$_2$,op,d,i]}}{\emptyset}{\True}}{\\\hspace*{20mm}\texttt{p.exists(hasUnevaluableExpr)}}
}
\formatpatternmatch{
\PatternNode{\texttt{Limit}}{[\texttt{c}]}{\\\hspace*{20mm}\PatternNode{\texttt{LocalRelation}}{\texttt{[o,d,i]}}{\emptyset}{\True}}{\True}
}
\formatpatternmatch{
\PatternNode{\texttt{Filter}}{\texttt{[o$_1$,r$_1$,c]}}{\\\hspace*{20mm}\PatternNode{\texttt{LocalRelation}}{\texttt{[o$_2$,r$_2$,op,d,i}]}{\emptyset}{\True}}{\\\hspace*{20mm}\texttt{c.exists(hasUnevaluableExpr)}}
}
\similartojitd{DeleteSingletonFromArray}

\subsection{\textbf{Transform:} CollapseProject}
\sparkscala
\begin{lstlisting}[style=scala]
case p1 : Project(_, p2: Project)
\end{lstlisting}
\begin{lstlisting}[style=scala]
case p : Project(_, agg: Aggregate)
\end{lstlisting}
\begin{lstlisting}[style=scala]
case Project(l1, g : GlobalLimit(_, limit : 
        LocalLimit(_, p2 : Project(l2, _))))
            if isRenaming(l1, l2)
\end{lstlisting}
\begin{lstlisting}[style=scala]
case Project(l1, limit : 
        LocalLimit(_, p2 : Project(l2, _)))
            if isRenaming(l1, l2)
\end{lstlisting}
\begin{lstlisting}[style=scala]
case Project(l1, limit : 
        LocalLimit(_, p2 : Project(l2, _)))
            if isRenaming(l1, l2)
\end{lstlisting}
\begin{lstlisting}[style=scala]
case Project(l1, r : 
        Repartition(_, _, p : Project(l2, _)))
            if isRenaming(l1, l2)
\end{lstlisting}
\begin{lstlisting}[style=scala]
case Project(l1, s : 
        Sample(_, _, _, _, p2 : Project(l2, _)))
            if isRenaming(l1, l2)
\end{lstlisting}
\begin{lstlisting}[style=scala]
private def isRenaming(list1: Seq[NamedExpression],
list2: Seq[NamedExpression]): Boolean = 
{
    list1.length == list2.length && 
    list1.zip(list2).forall {
      case (e1, e2) if e1.semanticEquals(e2) => true
      case (Alias(a: Attribute, _), b) 
      if a.metadata == Metadata.empty && a.name == 
      b.name => true
      case _ => false
    }
  }
}
\end{lstlisting}
\grammarpattern
\formatpatternmatch{
\PatternNode{\texttt{Project}}{\texttt{[o$_1$,r$_1$,p$_1$]}}{\\\hspace*{20mm}\PatternNode{\texttt{Project}}{\texttt{[o$_2$,r$_2$,p$_2$]}}{\Pattern_1}{\True}}{\True}
}
\formatpatternmatch{
\PatternNode{\texttt{Project}}{\texttt{[o$_1$,r$_1$,p]}}{\\\hspace*{20mm}\PatternNode{\texttt{Aggregate}}{\texttt{[o$_2$,r$_2$,g,a]}}{\Pattern_1}{\True}}{\True}
}
\formatpatternmatch{
\PatternNode{\texttt{Project}}{\texttt{[o$_1$,r$_1$,p$_1$]}}{\\\hspace*{20mm}\PatternNode{\texttt{Globallimit}}{\texttt{[o$_2$,r$_2$,g]}}{\\\hspace*{20mm}\PatternNode{\texttt{LocalLimit}}{\texttt{[o$_3$,r$_3$,l]}}{\\\hspace*{20mm}\PatternNode{\texttt{Project}}{\texttt{[o$_4$,r$_4$,p$_2$]}}{\Pattern_1}{\True}}{\True}}{\True}}{\\\hspace*{20mm}\texttt{isRenaming(p$_1$,p$_2$)}}
}
\formatpatternmatch{
\PatternNode{\texttt{Project}}{\texttt{[o$_1$,r$_1$,p$_1$]}}{\\\hspace*{20mm}\PatternNode{\texttt{LocalLimit}}{\texttt{[o$_2$,r$_2$,l]}}{\\\hspace*{20mm}\PatternNode{\texttt{Project}}{\texttt{[o$_3$,r$_3$,p$_1$]}}{\Pattern_1}{\True}}{\True}}{\\\hspace*{20mm}\texttt{isRenaming(p$_1$,p$_2$)}}
}
\formatpatternmatch{
\PatternNode{\texttt{Project}}{\texttt{[o$_1$,r$_1$,p$_1$]}}{\\\hspace*{20mm}\PatternNode{\texttt{Repartition}}{\texttt{[o$_2$,r$_2$,n,s]}}{\\\hspace*{20mm}\PatternNode{\texttt{Project}}{\texttt{[o$_3$,r$_3$,p$_2$]}}{\Pattern_1}{\True}}{\True}}{\\\hspace*{20mm}\texttt{isRenaming(p$_1$,p$_2$)}}
}

\formatpatternmatch{
\PatternNode{\texttt{Project}}{\texttt{[o$_1$,r$_1$,p$_1$]}}{\\\hspace*{20mm}\PatternNode{\texttt{Sample}}{\texttt{[o$_2$,r$_2$,l,u,w,s]}}{\\\hspace*{20mm}\PatternNode{\texttt{Project}}{\texttt{[o$_3$,r$_3$,p$_2$]}}{\Pattern_1}{\True}}{\True}}{\\\hspace*{20mm}\texttt{isRenaming(p$_1$,p$_2$)}}
}

\similartojitd{PivotLeft/Right,PushDownAndCrack}
The third match pattern represents a 4-way join which is a exception. Most other look at a 3-level deep subtree representative of a  3-way join.
\section{ORCA Transforms}
\label{sec: orca transforms}
Orca defines specific patterns for rewrites and matches at runtime the AST nodes. If the node matches the pattern a rewrite is defined over the node is replaced.  
Orca's AST grammar is slightly more expressive than our won in that certain \texttt{CExpression} nodes in ORCA like \texttt{CLogicalNaryJoin} support multiple children; however (i) this is a limitation we impose largely for simplicity of presentation, and (ii) none of the patterns we encountered include recursive patterns among the children of such variable-child nodes. 

In the following, we provide (i) the Pattern for rewrites along with the function that computes the rewrite's promise which determines the priority of the rewrite, (ii) The corresponding pattern matching expression in our grammar, and (iii) The most similar transform. Determining the priority of a rewrite is encoded in the grammar as a constraint over the pattern match.
\subsection{Transform: ExpandNaryJoin}
\orcacpp
\begin{lstlisting}[style=cpp]
CXformExpandNAryJoin::
CXformExpandNAryJoin(CMemoryPool *mp)
	: CXformExploration(
		  // pattern
		  GPOS_NEW(mp) CExpression(
			  mp, GPOS_NEW(mp) CLogicalNAryJoin(mp),
			  GPOS_NEW(mp) 
			  CExpression(mp, GPOS_NEW(mp) 
			  CPatternMultiLeaf(mp)),
			  GPOS_NEW(mp) 
			  CExpression(mp, GPOS_NEW(mp) 
			  CPatternTree(mp))))
{
}
CXform::EXformPromise
CXformExpandNAryJoin::
Exfp(CExpressionHandle &exprhdl) const
{
	if (exprhdl.DeriveHasSubquery(exprhdl.Arity() - 1))
	{
		// subqueries must be unnested before 
		// applying xform
		return CXform::ExfpNone;
	}
#ifdef GPOS_DEBUG
	CAutoMemoryPool amp;
	GPOS_ASSERT(!CXformUtils::
	FJoinPredOnSingleChild(amp.Pmp(), exprhdl) &&
				"join predicates are not pushed down");
#endif	// GPOS_DEBUG

	return CXform::ExfpHigh;
}
\end{lstlisting}
\grammarpattern
\formatpatternmatch{
\PatternNode{\texttt{CLogicalNAryJoin}}{\texttt{[exprhdl,p}]}{\PatternList}{\linebreak\hspace*{20mm}\texttt{exprhdl.hasSubQuery}}
}

\similartojitd{PushDownAndCrack}

\subsection{Transform: ExpandNaryJoinMinCard}
\orcacpp
\begin{lstlisting}[style=cpp]
CXformExpandNAryJoinMinCard::
CXformExpandNAryJoinMinCard(CMemoryPool *mp)
	: CXformExploration(
		  // pattern
		  GPOS_NEW(mp) CExpression(
			  mp, GPOS_NEW(mp) CLogicalNAryJoin(mp),
			  GPOS_NEW(mp) 
			  CExpression(mp, GPOS_NEW(mp)
			  CPatternMultiTree(mp)),
			  GPOS_NEW(mp) 
			  CExpression(mp, GPOS_NEW(mp)
			  CPatternTree(mp))))
{
}
CXform::EXformPromise
CXformExpandNAryJoinMinCard::
Exfp(CExpressionHandle &exprhdl) const
{
	return CXformUtils::
	ExfpExpandJoinOrder(exprhdl, this);
}
\end{lstlisting}
\grammarpattern
\formatpatternmatch{
\PatternNode{\texttt{CLogicalNAryJoin}}{\texttt{[exprhdl,p]}}{\PatternList}{\linebreak\hspace*{20mm}\texttt{exprhdl.expandJoinOrd}}
}

\similartojitd{PushDownAndCrack}

\subsection{Transform: ExpandNaryJoinGreedy}
\orcacpp
\begin{lstlisting}[style=cpp]
CXformExpandNAryJoinGreedy::
CXformExpandNAryJoinGreedy(CMemoryPool *pmp)
	: CXformExploration(
		  // pattern
		  GPOS_NEW(pmp) CExpression(
			  pmp, GPOS_NEW(pmp) 
			  CLogicalNAryJoin(pmp),
			  GPOS_NEW(pmp)
				  CExpression(pmp, GPOS_NEW(pmp)
				  CPatternMultiTree(pmp)),
			  GPOS_NEW(pmp) CExpression(pmp,
			  GPOS_NEW(pmp) CPatternTree(pmp))))
{
}
CXform::EXformPromise
CXformExpandNAryJoinGreedy::
Exfp(CExpressionHandle &exprhdl) const
{
	return CXformUtils::
	ExfpExpandJoinOrder(exprhdl, this);
}

\end{lstlisting}
\grammarpattern
\formatpatternmatch{
\PatternNode{\texttt{CLogicalNAryJoin}}{\texttt{[exprhdl,p}]}{\PatternList}{\linebreak\hspace*{20mm}\texttt{exprhdl.expandJoinOrd}}
}
\similartojitd{PushDownAndCrack}

\subsection{Transform: ExpandNAryJoinDP}
\orcacpp
\begin{lstlisting}[style=cpp]
CXformExpandNAryJoinDP::
CXformExpandNAryJoinDP(CMemoryPool *mp)
	: CXformExploration(
		  // pattern
		  GPOS_NEW(mp) CExpression(
			  mp, GPOS_NEW(mp) CLogicalNAryJoin(mp),
			  GPOS_NEW(mp) CExpression(mp, GPOS_NEW(mp)
			  CPatternMultiLeaf(mp)),
			  GPOS_NEW(mp) CExpression(mp, GPOS_NEW(mp)
			  CPatternTree(mp))))
{
}

CXform::EXformPromise
CXformExpandNAryJoinDP::
Exfp(CExpressionHandle &exprhdl) const
{
	COptimizerConfig *optimizer_config =
		COptCtxt::PoctxtFromTLS()->GetOptimizerConfig();
	const CHint *phint = optimizer_config->GetHint();

	const ULONG arity = exprhdl.Arity();

	// since the last child of the join operator is a 
	// scalar child
	// defining the join predicate, ignore it.
	const ULONG ulRelChild = arity - 1;

	if (ulRelChild > phint->UlJoinOrderDPLimit())
	{
		return CXform::ExfpNone;
	}

	return CXformUtils::
	ExfpExpandJoinOrder(exprhdl, this);
}
\end{lstlisting}
\grammarpattern
\formatpatternmatch{
\PatternNode{\texttt{CLogicalNAryJoin}}{\texttt{[exprhdl,p]}}{\PatternList}{\linebreak\texttt{exprhdl.Arity-1 > x || exprhdl.expandJoinOrd}}
}
\textbf{Code:}

\similartojitd{PushDownAndCrack}

\subsection{Transform: Get2TableScan}
\orcacpp
\begin{lstlisting}[style=cpp]
CXformGet2TableScan::CXformGet2TableScan(CMemoryPool *mp)
	: CXformImplementation(
		  // pattern
		  GPOS_NEW(mp) CExpression(mp, GPOS_NEW(mp)
		  CLogicalGet(mp)))
{
}
CXform::EXformPromise
CXformGet2TableScan::
Exfp(CExpressionHandle &exprhdl) const
{
	CLogicalGet *popGet =
	CLogicalGet::PopConvert(exprhdl.Pop());

	CTableDescriptor *ptabdesc = popGet->Ptabdesc();
	if (ptabdesc->IsPartitioned())
	{
		return CXform::ExfpNone;
	}

	return CXform::ExfpHigh;
}
\end{lstlisting}
\grammarpattern
\formatpatternmatch{
\PatternNode{\texttt{CLogicalGet}}{\texttt{[exprhdl,t,pt,o,k,c]}}{\emptyset}{\linebreak\hspace*{20mm}\texttt{pt.isPartitioned}}
}

\similartojitd{CrackArray}

\subsection{Transform: Select2Filter}
\orcacpp
\begin{lstlisting}[style=cpp]
CXformSelect2Filter::CXformSelect2Filter(CMemoryPool *mp)
	:  // pattern
	  CXformImplementation(GPOS_NEW(mp) CExpression(
		  mp, GPOS_NEW(mp) CLogicalSelect(mp),
		  GPOS_NEW(mp) CExpression(
			  mp, GPOS_NEW(mp) CPatternLeaf(mp)),  // relational child
		  GPOS_NEW(mp)
			  CExpression(mp, GPOS_NEW(mp) CPatternLeaf(mp))  // predicate
		  ))
{
}

CXform::EXformPromise
CXformSelect2Filter::Exfp(CExpressionHandle &exprhdl) const
{
	if (exprhdl.DeriveHasSubquery(1))
	{
		return CXform::ExfpNone;
	}

	return CXform::ExfpHigh;
}
\end{lstlisting}
\grammarpattern
\formatpatternmatch{
\PatternNode{\texttt{CLogicalSelect}}{\linebreak\hspace*{20mm}\texttt{[exprhdl,m,pt,PredicateExpr]}}{\Pattern_1}{\linebreak\hspace*{20mm}\texttt{exprhdl.hasSubQuery}}
}

\similartojitd{CrackArray}

\subsection{Transform: InnerJoin2NLJoin}
\orcacpp
\begin{lstlisting}[style=cpp]
CXformInnerJoin2NLJoin::
CXformInnerJoin2NLJoin(CMemoryPool *mp)
	:  // pattern
	  CXformImplementation(GPOS_NEW(mp) CExpression(
		  mp, GPOS_NEW(mp) CLogicalInnerJoin(mp),
		  GPOS_NEW(mp)
			  CExpression(mp, GPOS_NEW(mp)
			  CPatternLeaf(mp)),  // left child
		  GPOS_NEW(mp)
			  CExpression(mp, GPOS_NEW(mp)
			  CPatternLeaf(mp)),  // right child
		  GPOS_NEW(mp)
			  CExpression(mp, GPOS_NEW(mp)
			  CPatternLeaf(mp))  // predicate
		  ))
{
}
CXform::EXformPromise
CXformInnerJoin2NLJoin::
Exfp(CExpressionHandle &exprhdl) const
{
	return CXformUtils::
	ExfpLogicalJoin2PhysicalJoin(exprhdl);
}
\end{lstlisting}
\grammarpattern
\formatpatternmatch{
\PatternNode{\texttt{CLogicalInnerJoin}}{\\\hspace*{20mm}\texttt{[exprhdl,predicateExpr]}}{\Pattern_1,\Pattern_2}{\linebreak\hspace*{10mm}\texttt{exprhdl.ExfpLogicalJoin2PhysicalJoin}}
}

\similartojitd{PushDownAndCrack} 

\subsection{Transform: InnerJoin2HashJoin}
\orcacpp
\begin{lstlisting}[style=cpp]
CXformInnerJoin2NLJoin::
CXformInnerJoin2NLJoin(CMemoryPool *mp)
	:  // pattern
	  CXformImplementation(GPOS_NEW(mp) CExpression(
		  mp, GPOS_NEW(mp) CLogicalInnerJoin(mp),
		  GPOS_NEW(mp)
			  CExpression(mp, GPOS_NEW(mp)
			  CPatternLeaf(mp)),  // left child
		  GPOS_NEW(mp)
			  CExpression(mp, GPOS_NEW(mp)
			  CPatternLeaf(mp)),  // right child
		  GPOS_NEW(mp)
			  CExpression(mp, GPOS_NEW(mp)
			  CPatternLeaf(mp))  // predicate
		  ))
{
}
CXform::EXformPromise
CXformInnerJoin2NLJoin::
Exfp(CExpressionHandle &exprhdl) const
{
	return CXformUtils::
	ExfpLogicalJoin2PhysicalJoin(exprhdl);
}
\end{lstlisting}
\grammarpattern
\formatpatternmatch{
\PatternNode{\texttt{CLogicalInnerJoin}}{\\\hspace*{10mm}\texttt{[exprhdl,predicateExpr]}}{\Pattern_1,\Pattern_2}{\\\hspace*{10mm}\texttt{exprhdl.ExfpLogicalJoin2PhysicalJoin}}
}

\similartojitd{PushDownAndCrack}

\subsection{Transform: JoinCommutativity}
\orcacpp
\begin{lstlisting}[style=cpp]
CXformJoinCommutativity::
CXformJoinCommutativity(CMemoryPool *mp)
	: CXformExploration(
		  // pattern
		  GPOS_NEW(mp) CExpression(
			  mp, GPOS_NEW(mp) CLogicalInnerJoin(mp),
			  GPOS_NEW(mp)
				  CExpression(mp, GPOS_NEW(mp)
				  CPatternLeaf(mp)),  // left child
			  GPOS_NEW(mp) CExpression(
				  mp, GPOS_NEW(mp) 
				  CPatternLeaf(mp)),  // right child
			  GPOS_NEW(mp)
				  CExpression(mp, GPOS_NEW(mp)
				  CPatternLeaf(mp)))  // predicate
	  )
{
}

BOOL
CXformJoinCommutativity::FCompatible(CXform::EXformId exfid)
{
	BOOL fCompatible = true;

	switch (exfid)
	{
		case CXform::ExfJoinCommutativity:
			fCompatible = false;
			break;
		default:
			fCompatible = true;
	}

	return fCompatible;
}
\end{lstlisting}
\grammarpattern
\formatpatternmatch{
\PatternNode{\texttt{CLogicalInnerJoin}}{\texttt{[exprhdl,predicateExpr]}}{\\\hspace*{20mm}\Pattern_1,\Pattern_2}{\texttt{exprhdl.id}}
}
\similartojitd{PivotLeft/Right}

\subsection{Transform: ImplementUnionAll}
\orcacpp
\begin{lstlisting}[style=cpp]
CXformImplementUnionAll::
CXformImplementUnionAll(CMemoryPool *mp)
	:  // pattern
	  CXformImplementation(GPOS_NEW(mp) 
	  CExpression(
		  mp, GPOS_NEW(mp) CLogicalUnionAll(mp),
		  GPOS_NEW(mp) CExpression(mp, GPOS_NEW(mp)
		  CPatternMultiLeaf(mp))))
{
}
\end{lstlisting}
\grammarpattern
\formatpatternmatch{
\PatternNode{\texttt{CLogicalUnionAll}}{\texttt{[exprhdl,i,o,k]}}{\PatternList}{\True}
}

\similartojitd{MergeUnsorted/SortedConcat}

}{}
\end{document}